\documentclass[aps,prd,preprint,superscriptaddress,tighten,nofootinbib]{revtex4}
\usepackage{amssymb}
\usepackage{mathrsfs}
\usepackage{graphicx}
\usepackage{amsmath}
\usepackage{epsfig}
\usepackage{color}
\def\slc#1{\setbox0=\hbox{$#1$}           % set a box for #1
    \dimen0=\wd0                                 % and get its size
    \setbox1=\hbox{/} \dimen1=\wd1               % get size of /
    \ifdim\dimen0>\dimen1                        % #1 is bigger
       \rlap{\hbox to \dimen0{\hfil/\hfil}}      % so center / in box
       #1                                        % and print #1
    \else                                        % / is bigger
       \rlap{\hbox to \dimen1{\hfil$#1$\hfil}}   % so center #1
       /                                         % and print /
    \fi}

\begin{document}

\draft
\preprint{RM3-TH/09-1}

\title{Exact and Approximate Formulas for Neutrino Mixing and Oscillations with Non-Standard Interactions}

\author{Davide Meloni}
\email{meloni@fis.uniroma3.it}

\affiliation{Dipartimento di Fisica, Universit{\'a} di Roma Tre and
INFN Sez.~di Roma Tre, via della Vasca Navale 84, 00146 Roma, Italy}

\author{Tommy Ohlsson}
\email{tommy@theophys.kth.se}

\author{He Zhang}
\email{zhanghe@kth.se}

\affiliation{Department of Theoretical Physics, School of
Engineering Sciences, Royal Institute of Technology (KTH) --
AlbaNova University Center, Roslagstullsbacken 21, 106 91 Stockholm,
Sweden}

\begin{abstract}
We present, both exactly and approximately, a complete set of mappings
between the vacuum (or fundamental) leptonic mixing parameters and the
effective ones in matter with non-standard neutrino interaction (NSI)
effects included. Within the three-flavor neutrino framework and a
constant matter density profile, a full set of sum rules is
established, which enables us to reconstruct the moduli of the
effective leptonic mixing matrix elements, in terms of the vacuum
mixing parameters in order to reproduce the neutrino oscillation
probabilities for future long-baseline experiments. Very compact, but
quite accurate, approximate mappings are obtained based on series
expansions in the neutrino mass hierarchy parameter $\eta \equiv
\Delta m^2_{21}/\Delta m^2_{31}$, the vacuum leptonic mixing parameter
$s_{13} \equiv \sin\theta_{13}$, and the NSI parameters
$\varepsilon_{\alpha\beta}$. A detailed numerical analysis about how
the NSIs affect the smallest leptonic mixing angle $\theta_{13}$, the
deviation of the leptonic mixing angle $\theta_{23}$ from its maximal
mixing value, and the transition probabilities useful for future
experiments are performed using our analytical results.
\end{abstract}

\maketitle

\section{Introduction}
\label{sec:intro}

During the past decade, neutrino oscillation experiments have
provided us with very convincing evidence that neutrinos are massive
and lepton flavors are mixed
\cite{Fukuda:1998mi,Fukuda:2001nj,cc:2008zn,Ahmad:2001an,
Ahmed:2003kj,Aharmim:2008kc,Apollonio:1999ae,Eguchi:2002dm,Ahn:2002up,
Michael:2006rx}. It opens an important window for searching new
physics beyond the Standard Model (SM) of particle physics, and has
significant cosmological implications. Within the framework of three
active neutrinos, neutrino masses are the leading mechanism behind
neutrino oscillations
\cite{Pontecorvo:1957cp,Maki:1962mu,Pontecorvo:1967fh,Gribov:1968kq}.
However, in future long-baseline neutrino oscillation experiments,
besides the standard matter effects
\cite{Wolfenstein:1977ue,Mikheev:1986gs}, the possibility of testing
non-standard neutrino interactions (NSIs) should be opened up.

Note that, NSIs enter neutrino oscillations at production,
propagation, and detection processes. In principle, in the case of
dimension-6 operators, the corresponding NSI parameters are related
to the underlying new physics in the form of $\varepsilon \sim
(m_W/m_X)^2$, where $m_W$ is the mass of the W boson and $m_X$
denotes the new physics energy scale. A rough estimate indicates
that if new physics appears at the TeV region, the magnitude of NSI
parameters should not be larger than a few percent, although the
present experimental upper bounds are still very loose. Due to the
interference effects, NSIs modify the standard flavor transitions at
leading order in $\varepsilon$ for some typical processes,
especially at a future neutrino factory or other facilities with
high-energy beams
\cite{GonzalezGarcia:2001mp,Gago:2001xg,Huber:2001zw,Ota:2001pw,NSInfstart,NSInfstop,Blennow:2004js,Blennow:2005qj,NSIotherstart,NSIotherstop,Adhikari:2006uj,Blennow:2007pu,Kopp:2007mi,
Ribeiro:2007ud,Kopp:2007ne,Kopp:2008ds,Blennow:2008ym,Winter:2008eg,Altarelli:2008yr,Kikuchi:2008vq}.
In these cases, NSI corrections become particularly relevant, and
the experimentally measured values of leptonic mixing angles and CP
violating quantities are dramatically different from the vacuum
ones. In this sense, the combination of standard neutrino
oscillations and NSI effects in the analyses of future neutrino
experiments is not only meaningful but also necessary. In addition
to NSIs at propagation processes, NSI effects at neutrino sources
and detectors play a very important role, since they may induce
significant mimicking effects on mixing parameters
\cite{Ohlsson:2008gx} or bring in distinguishable zero-distance
effects for a near detector \cite{Bueno:2000jy,Malinsky:2008qn}.
Here we will only concentrate on NSIs during the phase of neutrino
propagation, and a brief discussion on how to consistently include
the source and detector effects will be implemented at the end of
Sec.~\ref{sec:effective}.

Although much attention has been paid on the issue of NSIs according
to different neutrino facilities and projects, the previous
analytical investigations are either based on a two-flavor neutrino
framework \cite{Blennow:2008eb} or an approximation for the
three-flavor neutrino oscillation probabilities (indeed producing
lengthy formulas). There are still lack of analytical relations,
which can show us the NSI effects on the leptonic mixing parameters
in a transparent way. Thus, in this paper, we first develop a full
set of sum rules, which relate the vacuum leptonic mixing matrix
elements and their effective counterparts in matter. By solving
these sum rules, it is straightforward to establish the leptonic
mixing matrix, unitarity triangles \cite{Farzan:2002ct}, and CP
violating effects in matter (see Sec.~\ref{sec:sum_rules}). We then
present series expansions of mappings in the mass hierarchy
parameter $\eta \equiv \Delta m^2_{21}/\Delta m^2_{31}$, the mixing
parameter $s_{13} \equiv \sin\theta_{13}$, and the NSI parameters
$\varepsilon_{\alpha\beta}$. The NSI corrections to the vacuum
mixing parameters can be manifested in a distinct way. We hope that
the elegant and compact formulas provided in this paper could be
very helpful for the phenomenological studies of future
long-baseline neutrino oscillation experiments.

This work is organized as follows. In Sec.~\ref{sec:effective}, we
present the general formulas and notations, and show how the
neutrino oscillation probabilities can be expressed through the
language of effective mixing parameters. In
Sec.~\ref{sec:sum_rules}, we introduce the sum rules between
leptonic mixing matrix elements and the effective counterparts in
matter, and then derive the mappings exactly. The expressions of
effective masses in matter, which are necessary for the calculation
of neutrino oscillation probabilities, are shown in detail in
Appendix~\ref{app:effective_masses}. Next, in
Sec.~\ref{sec:expansions}, we derive a full set of series expansions
of these mappings. We also compare our mapping results with the
corresponding expressions existing in the literature but without
NSIs, and find that our results are in agreement with previous
analyses in the limit of vanishing NSIs.
Section~\ref{sec:applications} is devoted to applications of our
analytical mappings. Numerical illustrations in order to show the
validity and reliability of our approximate results are also
presented. Finally, a brief summary is given in
Sec.~\ref{sec:summary}.

\section{The language of effective parameters}
\label{sec:effective}

At energy scales $\mu\ll m_W$, the NSIs involving neutrinos can be
described by the effective Lagrangian
\begin{eqnarray}\label{eq:L}
{\cal L}_{\rm NSI} = -\frac{G_F}{\sqrt{2}}
\sum_{f,P}\varepsilon^{fP}_{\alpha\beta} \left(
\overline{\nu_\alpha}\gamma^\mu L \nu_\beta \right) \left(
\overline{f}\gamma_\mu P f \right) \ ,
\end{eqnarray}
where $f$ is a charged lepton or quark, $G_F$ is the Fermi coupling
constant, and $P=\{ L, R\}$ is a projection operator. The parameters
$\varepsilon^{fP}_{\alpha\beta}$, which are entries of the Hermitian
matrix $\varepsilon^{fP}$, give the strengths of the NSIs. The
magnitudes of the NSI parameters can be constrained from neutrino
deep inelastic scattering experiments and from elastic $\nu-e$
scattering in which the NSIs would contribute to the determination
of $\sin^2\theta_{W}$, i.e., the Weinberg angle. The latest
constraints on $\varepsilon^{fP}_{\alpha\beta}$ have been discussed
in
Ref.~\cite{Berezhiani:2001rs,Davidson:2003ha,Barranco:2005ps,Barranco:2007ej},
and the most stringent bounds are those on
$\varepsilon^{fP}_{\mu\alpha}$ for $\alpha=e,\mu,\tau$.

In order to introduce the effective mixing parameters, we start from
neutrino oscillations in vacuum. The evolution in time of a neutrino
state $|\nu(t)\rangle$ is given by the Schr{\"o}dinger-like equation
\begin{eqnarray}\label{eq:schordinger}
{\rm i} \frac{\rm d}{{\rm d}t} | \nu(t) \rangle = H | \nu(t) \rangle
\ ,
\end{eqnarray}
where $H$ is the Hamiltonian of the system. For neutrinos traveling
in vacuum, the Hamiltonian in the ultra-relativistic limit $E\gg
m_{i}$ reads
\begin{eqnarray}\label{eq:Hvacuum}
H = \frac{1}{2E} V {\rm diag}\left( 0,\Delta_{21},\Delta_{31}
\right) V^{\dagger} \ ,
\end{eqnarray}
where $\Delta_{ij} \equiv m^2_i -m ^2_j$ are the neutrino
mass-squared differences and $E$ denotes the neutrino energy. In
addition, $V$ is the unitary leptonic mixing matrix
\cite{Maki:1962mu}, which relates the mass eigenstates of the three
neutrinos $(\nu_1, \nu_2, \nu_3)$ to their corresponding flavor
eigenstates $(\nu_e, \nu_\mu, \nu_\tau)$
\begin{eqnarray}\label{eq:V}
\nu_\alpha = \sum_i V_{\alpha i} \nu_i \; ,
\end{eqnarray}
for $\alpha=e,\mu,\tau$. For simplicity, the sum of Latin indices
run over $1,2,3$ and the sum of Greek indices run over $e,\mu,\tau$
throughout this paper, if not otherwise stated. We can define the
evolution matrix $S(t,t_0)$ such that
\begin{eqnarray}\label{eq:S}
| \nu(t) \rangle = S(t,t_0)  | \nu(t_0) \rangle \ , \ \ \ \ \
S(t_0,t_0) = 1 \ ,
\end{eqnarray}
and $S(t,t_0)$ satisfies the same Schr{\"o}dinger-like
equation~\eqref{eq:schordinger}, as $| \nu(t) \rangle$. The neutrino
oscillation probabilities can be found as
$P_{\alpha\beta}=|S_{\beta\alpha}(t,t_0)|^2$. Using
Eq.~\eqref{eq:Hvacuum}, the elements of the evolution matrix are
given by
\begin{eqnarray}\label{eq:Ssolution}
S_{\beta\alpha}(t,t_0) = \sum_i V_{\alpha i} V^*_{\beta i} e^{-{\rm
i}\frac{ m^2_i L}{2E} } \ ,
\end{eqnarray}
where we have identified $L\equiv t-t_0$. Thus, the probability of
transition from a neutrino flavor $\alpha$ to a neutrino flavor
$\beta$ is given by
\begin{eqnarray}\label{eq:P}
P_{\alpha\beta} \equiv \left|S_{\beta\alpha}(t,t_0)\right|^2 =
\left|  \sum_i V_{\alpha i} V^*_{\beta i} e^{-{\rm i}\frac{ m^2_i
L}{2E} } \right|^2 \ .
\end{eqnarray}

For the setups of future long-baseline neutrino oscillation
experiments, the neutrino beams inevitably travel through the
Earth's mantle, and the charged-current contributions to the
matter-induced effective potential have to be considered properly.
Disregarding the neutral-current contributions, the effective
Hamiltonian responsible for neutrino propagation in matter is given
by
\begin{eqnarray}\label{eq:Hm}
\tilde{H}_{\alpha\beta} = H_{\alpha\beta} + a\left( \delta_{\alpha
e} \delta_{\beta e} + \varepsilon_{\alpha\beta} \right) \ ,
\end{eqnarray}
where the matter parameter $a=\sqrt{2}G_F N_e $ arises from
coherent forward scattering. Here $N_e$ denotes the electron number
density along the neutrino trajectory in the Earth and the NSI
parameters $\varepsilon_{\alpha\beta}$ are defined as
\begin{eqnarray}\label{eq:eps}
\varepsilon_{\alpha\beta} = \sum_{f,P}
\varepsilon^{fP}_{\alpha\beta} \frac{N_f}{N_e} \ ,
\end{eqnarray}
with $N_f$ being the number density of a fermion of type $f$.

Similar to the vacuum Hamiltonian in Eq.~\eqref{eq:Hvacuum}, the
effective Hamiltonian in matter can also be diagonalized through a
unitary transformation
\begin{eqnarray}\label{eq:Heff}
\tilde{H} = \frac{1}{2E} \tilde{V} {\rm diag}\left(
\tilde{m}^2_1,\tilde{m}^2_2, \tilde{m}^2_3 \right)
\tilde{V}^{\dagger} \ ,
\end{eqnarray}
where $\tilde{m}^{2}_i$ denote the effective mass-squared
eigenvalues of neutrinos and $\tilde V$ is the unitary mixing matrix
in matter. Note that, in writing down Eq.~\eqref{eq:Heff}, we have
already taken into account the Hermitian property of $\tilde{H}$.

In order to write out explicitly the transition probabilities, we
assume a constant matter density profile, which is actually close to
reality for most of the proposed long-baseline experiments. Following
analogous procedures as shown in Eqs.~(\ref{eq:V})-(\ref{eq:P}), one
can then obtain the transition probabilities with matter effects
included as
\begin{eqnarray}\label{eq:Pm}
P_{\alpha\beta} \equiv \left|S_{\beta\alpha}(t,t_0)\right|^2 =
\left|  \sum_i \tilde V_{\alpha i} \tilde V^*_{\beta i} e^{-{\rm
i}\frac{ \tilde m^2_i L}{2E} } \right|^2 \ .
\end{eqnarray}
Comparing Eq.~\eqref{eq:P} with Eq.~\eqref{eq:Pm}, we arrive at the
conclusion that there is no difference between the form of the
neutrino oscillation probabilities in vacuum and in matter if we
replace the vacuum parameters $V$ and $m_i^2$ by the effective
parameters $\tilde V$ and $\tilde m^2_i$. The mappings between
vacuum parameters and the effective ones are sufficient in order to
study the new physics effects entering future long-baseline neutrino
oscillation experiments. The key point turns out to be the
diagonalization of the effective Hamiltonian $\tilde H$ and figuring
out the explicit relations of the effective parameters.

In most of the viable models for NSIs, the source and detector
effects are simultaneously taken into account, despite their
magnitude. The language of effective mixing parameters can then
easily be extended to the case including NSIs at neutrino sources
and detectors. Now, the NSI parameters at sources and detectors can
be defined as
\cite{Grossman:1995yp,GonzalezGarcia:2001mp,Kopp:2007ne}
\begin{eqnarray}\label{eq:SD}
|\nu^s_\alpha \rangle & = &|\nu_\alpha \rangle +
\sum_{\beta=e,\mu,\tau} \varepsilon^s_{\alpha\beta}
|\nu_\beta\rangle   \ , \\ \langle \nu^d_\beta| & = &  \langle
\nu_\beta | + \sum_{\alpha=e,\mu,\tau} \varepsilon^d_{\alpha \beta}
\langle  \nu_\alpha  | \ ,
\end{eqnarray}
where the superscripts `$s$' and `$d$' denote source and detector,
respectively. The transition probabilities are then modified
as\footnote{Here we have neglected the normalization factors, which
are needed in order to normalize the quantum states.}
%%%%%%%%%%%%%%%%%%%%%%%%%%%%%%%%%%%%%%%
\begin{eqnarray}\label{eq:Ps}
P_{\alpha\beta} & = & \left| \left[\left( 1 + \varepsilon^d
\right)^T \cdot S(t,t_0) \cdot \left( 1 + \varepsilon^s \right)^T
\right]_{\beta\alpha} \right|^2 \nonumber \\
& = & \left| \sum_{\gamma,\delta,i} \left( 1 + \varepsilon^d
\right)_{\gamma\beta}  \left( 1 + \varepsilon^s
\right)_{\alpha\delta} \tilde V_{\delta i} \tilde V^*_{\gamma i}
e^{-{\rm i}\frac{ \tilde m^2_i L}{2E} } \right|^2 \ .
\end{eqnarray}
Note that Eq.~\eqref{eq:Ps} is also suitable to describe neutrino
oscillations with a non-unitary mixing matrix, i.e., the minimal
unitarity violation model \cite{Antusch:2006vwa}. In the following
sections, we will only concentrate on NSI effects during propagation
processes and establish parameter mappings both exactly and
approximately.

\section{Sum rules and parameter mappings}
\label{sec:sum_rules}

In order to establish analytical relations between the matrix
elements of $\tilde{V}$ and those of $V$, we develop a set of sum
rules, which {enables} us to express the products $\tilde{V}_{\alpha
i } \tilde{V}^{*}_{\beta i } $ by using $V$, $m^2_i$ and $\tilde
m^2_i$. Such an approach has partially been employed in
Refs.~\cite{Zhang:2004hf,Xing:2005gk,Zhang:2006yq} in the case of
three or four-neutrino mixing. Here we will work out the most
general form with both the standard matter effects and the NSI
effects included.

The first sum rule is just the unitarity conditions, which hold for
both $\tilde{V}$ and $V$,
\begin{eqnarray}\label{eq:sum1}
\sum_{i} \tilde{V}_{\alpha i} \tilde{V}^{*}_{\beta i} = \sum_{i}
{V}_{\alpha i} {V}^{*}_{\beta i} = \delta_{\alpha \beta} \ .
\end{eqnarray}
Inserting Eqs.~\eqref{eq:Hvacuum} and \eqref{eq:Heff} into
Eq.~\eqref{eq:Hm}, and comparing both sides of this result, it is
straightforward to obtain the second sum rule
\begin{eqnarray}\label{eq:sum2}
\sum_{i} \tilde{m}^2_i \tilde{V}_{\alpha i} \tilde{V}^{*}_{\beta i}
= \sum_{i} \Delta_{i1} {V}_{\alpha i} {V}^{*}_{\beta i} + {\cal
A}_{\alpha\beta} \ ,
\end{eqnarray}
where we have defined ${\cal A}_{\alpha\beta} \equiv A
\left(\delta_{\alpha e } \delta_{\beta e}+\varepsilon_{\alpha
\beta}\right)$ with $A \equiv 2Ea$ for simplicity. In order to
derive a linearly independent sum rule besides the first two, we
square both sides of Eq.~\eqref{eq:sum2} and obtain the squared
relation
\begin{eqnarray}\label{eq:sum3}
\sum_{i} \tilde{m}^4_i \tilde{V}_{\alpha i} \tilde{V}^{*}_{\beta i}
= \sum_{i} \Delta^2_{i1} {V}_{\alpha i} {V}^{*}_{\beta i} +
\sum_{\gamma} {\cal A}_{\alpha\gamma} {\cal A}^{*}_{\beta\gamma} +
\sum_{\gamma,i} \Delta_{i1} \left( {\cal A}_{\alpha\gamma}
{V}_{\gamma i} {V}^{*}_{\beta i} +{\cal A}^{*}_{\beta\gamma}
{V}_{\alpha i} {V}^{*}_{\gamma i}  \right) \ .
\end{eqnarray}
Equations~(\ref{eq:sum2})-(\ref{eq:sum3}) together with the unitarity
condition Eq.~\eqref{eq:sum1} construct a full set of linear equations
of $\tilde{V}_{\alpha i} \tilde{V}^{*}_{\beta i}$ (for $i=1,2,3$). By
solving this set of equations, one will arrive at the explicit
expressions of $\tilde{V}_{\alpha i} \tilde{V}^{*}_{\beta i}$
straightforwardly.

In order to be concrete, we reexpress those equations in the
following form
\begin{eqnarray}\label{eq:linear}
\tilde{O}\left(\begin{matrix}\tilde{V}_{\alpha 1}
\tilde{V}^{*}_{\beta 1} \cr \tilde{V}_{\alpha 2}
\tilde{V}^{*}_{\beta 2} \cr \tilde{V}_{\alpha 3}
\tilde{V}^{*}_{\beta 3}\end{matrix}\right) & = & O
\left(\begin{matrix}{V}_{\alpha 1} {V}^{*}_{\beta 1} \cr {V}_{\alpha
2} {V}^{*}_{\beta 2} \cr {V}_{\alpha 3} {V}^{*}_{\beta
3}\end{matrix}\right)  + \left(\begin{matrix} 0 \cr {\cal
A}_{\alpha\beta} \cr  \sum_{\gamma} {\cal A}_{\alpha\gamma} {\cal
A}^*_{\beta\gamma}
\end{matrix}\right) \nonumber \\ & + & Q \sum_{\gamma} \left[ {\cal
A}_{\alpha\gamma} \left(\begin{matrix}  {V}_{\gamma 1}
{V}^{*}_{\beta 1} \cr {V}_{\gamma 2} {V}^{*}_{\beta 2} \cr
{V}_{\gamma 3} {V}^{*}_{\beta 3}
\end{matrix}\right) +{\cal
A}^{*}_{\beta\gamma} \left(\begin{matrix}  {V}_{\alpha 1}
{V}^{*}_{\gamma 1} \cr {V}_{\alpha 2} {V}^{*}_{\gamma 2} \cr
{V}_{\alpha 3} {V}^{*}_{\gamma 3}
\end{matrix}\right) \right] \ ,
\end{eqnarray}
where the matrices $\tilde{O},O$, and $Q$ are defined by
\begin{eqnarray}\label{eq:O}
\tilde{O}= \left(\begin{matrix} 1 & 1 & 1 \cr \tilde{m}^2_1 &
\tilde{m}^2_2 & \tilde{m}^2_3 \cr \tilde{m}^4_1 & \tilde{m}^4_2 &
\tilde{m}^4_3 \end{matrix}\right) \ , \ \ \ \ \ {O}=
\left(\begin{matrix} 1 & 1 & 1 \cr 0 & \Delta_{21} & \Delta_{31} \cr
0 & \Delta^{2}_{21} & \Delta^{2}_{31}
\end{matrix}\right) \ , \ \ \ \ \ {Q}=
\left(\begin{matrix} 0 & 0 & 0 \cr 0 & 0 & 0  \cr 0 & \Delta_{21} &
\Delta_{31}
\end{matrix}\right) \ .
\end{eqnarray}
After a lengthy calculation, the solution of Eq.~\eqref{eq:linear} can
be presented in a very elegant and compact form
\begin{eqnarray}\label{eq:solution}
\tilde{V}_{\alpha i} \tilde{V}^{*}_{\beta i} & = &
\frac{1}{\tilde{\Delta}_{im}\tilde{\Delta}_{in}} \left[ \sum_{j}
\hat{\Delta}_{jm} \hat{\Delta}_{jn} {V}_{\alpha j} {V}^{*}_{\beta j}
- {\cal A}_{\alpha\beta} \left(\tilde{m}^2_n
+\tilde{m}^2_m \right) \right. \nonumber \\
&& \left.  +\sum_{\gamma} {\cal A}_{\alpha\gamma} {\cal
A}^{*}_{\beta\gamma} + \sum_{\gamma,j} {\Delta}_{j1} \left( {\cal
A}_{\alpha\gamma}  {V}_{\gamma j} {V}^{*}_{\beta j}  + {\cal
A}^{*}_{\beta\gamma}  {V}_{\alpha j} {V}^{*}_{\gamma j} \right)
\right] \ ,
\end{eqnarray}
where $i\neq n \neq m$, $\tilde{\Delta}_{ij} =\tilde{m}^2_i -
\tilde{m}^2_j $, and $\hat{\Delta}_{ij} ={m}^2_i -m^2_1
-\tilde{m}^2_j $. Note that the pairs $(m,n)=(2,3)$, $(3,1)$,
$(1,2)$ in the right-hand side correspond to $i=1,2,3$ in the
left-hand side, respectively. Equation~(\ref{eq:solution}) is our
main result for the exact analytical mappings.

The full mappings require the explicit form of $\tilde m^2_i$, which
involves the cubic roots of the characteristic polynomial of
Eq.~\eqref{eq:Hm}. We follow the method given in
Ref.~\cite{Kopp:2008}, and the corresponding roots (or eigenvalues)
can be found in Appendix~\ref{app:effective_masses}. One may worry
about the $\tilde{m}^2_i$'s appearing in Eq.~\eqref{eq:solution},
since it seems that the effective mixing matrix elements rely on the
absolute effective neutrino masses.  However, recalling the
solutions in Eqs.~\eqref{eq:A4}-\eqref{eq:A6}, it is easy to observe
that only the mass-squared differences enter into the expressions of
$\tilde{m}^2_i$, and it guarantees the consistency of our
calculations. Obviously, ${\cal A}_{\alpha\beta} = 0$ leads to
$\tilde{V}_{\alpha i} \tilde{V}^{*}_{\beta i} = {V}_{\alpha i}
{V}^{*}_{\beta i}$. In the limit $\varepsilon_{\alpha\beta}
\rightarrow 0$, Eq.~\eqref{eq:solution} reduces to the case of
standard matter effects, and the main results given in
Refs.~\cite{Kimura:2002hb,Kimura:2002wd,Zhang:2004hf,Xing:2005gk,Yasuda:2007jp}
will be easily reproduced. These exact relations are model
independent and do not rely on any specific parametrization, and
hence, it will be very helpful to systematically study NSIs in
future experiments.

Taking $\alpha=\beta$, the moduli of $\tilde{V}_{\alpha i}$ can be
estimated immediately. For the case $\alpha \neq \beta$, the sides of leptonic
unitarity triangles, which are defined by the orthogonality
relations in Eq.~\eqref{eq:sum1} in the complex plane, are obtained.
These unitarity triangles have 18 different sides and nine different
inner angles, but their areas are all identical to a single
rephasing-invariant parameter ${\cal J}/2$ defined through
\cite{Jarlskog:1985ht}
\begin{eqnarray}\label{eq:J}
{\rm Im} ( V_{\alpha i} V_{\beta j} V^*_{\alpha j} V^*_{\beta i} ) =
{\cal J} \sum_{\gamma, k} ( \epsilon_{\alpha\beta\gamma}
\epsilon_{ijk} ) \ .
\end{eqnarray}
One of the major challenges of future long-baseline neutrino
oscillation experiments is to measure $\cal J$, in order to
establish the existence of CP violation in the lepton sector. We can
also define the counterpart of $\cal J$ in matter as $\tilde{\cal
J}$. Its magnitude is related to the moduli of the effective mixing
matrix elements as
\begin{eqnarray}\label{eq:Jm}
\tilde{\cal J}^2 & = & |\tilde{V}_{\alpha i}|^2 |\tilde{V}_{\beta
j}|^2 |\tilde{V}_{\alpha j}|^2 |\tilde{V}_{\beta i} |^2 -
\frac{1}{4} \left (1 + |\tilde{V}_{\alpha i}|^2 |\tilde{V}_{\beta
j}|^2 + |\tilde{V}_{\alpha j}|^2 |\tilde{V}_{\beta i}|^2 \right .
\nonumber \\
& & \left . - |\tilde{V}_{\alpha i}|^2 - |\tilde{V}_{\beta j}|^2 -
|\tilde{V}_{\alpha j}|^2 - |\tilde{V}_{\beta i}|^2 \right )^2 \; .%
\end{eqnarray}
As an application, we show, in Appendix~\ref{app:matrix elements}, the
zeroth-order series expansions of $|\tilde V_{e3}|^2$, $|\tilde
V_{e2}|^2$, and $|\tilde V_{\mu3}|^2$ in small parameters, i.e.,
$\eta$ and $V_{e3}$. In addition, simplified formulas of
Eq.~\eqref{eq:solution} and the effective mixing matrix elements for
standard matter effects are presented in Appendix~\ref{app:matrix
elements}.

Now, the neutrino oscillation probabilities can be directly obtained
with the help of Eq.~\eqref{eq:Pm} for a realistic experiment. In
order to be explicit, we can express the neutrino oscillation
probabilities in matter as
\begin{eqnarray}\label{eq:PJ1}
P_{\alpha\alpha} & = & 1-4\sum_{i>j} |\tilde V_{\alpha i} \tilde
V^*_{\alpha j}|^2 \sin^2\left(\frac{\tilde \Delta_{ij}L}{4E}\right) \ , \\
P_{\alpha\beta} & = &  - 4\sum_{i>j} {\rm Re}\left( \tilde
V^*_{\alpha i} \tilde V_{\beta i}\tilde V_{\alpha j}\tilde
V^*_{\beta j}\right)\sin^2\left(\frac{\tilde
\Delta_{ij}L}{4E}\right) - 8 \tilde {\cal J} \prod_{i>j}
\sin\left(\frac{\tilde \Delta_{ij}L}{4E}\right) \ , \label{eq:PJ2}
\end{eqnarray}
where $(\alpha,\beta)$ run over $(e,\mu)$, $(\mu,\tau)$, and
$(\tau,e)$. For anti-neutrino propagation in matter, we can simply
recalculate Eqs.~\eqref{eq:PJ1} and \eqref{eq:PJ2} through the
replacements $ A\to -A $, $V_{\alpha i} \to V^*_{\alpha i}$, and
$\varepsilon_{\alpha\beta} \to \varepsilon^*_{\alpha\beta}$. In
general, note that $\tilde V_{\alpha i}$, $\tilde \Delta_{ij}$ and
$\tilde {\cal J}$ for neutrinos are not identical to $\tilde
V_{\alpha i}$, $\tilde \Delta_{ij}$ and $\tilde {\cal J}$ for
anti-neutrinos. At first glance, one may wonder if the information
on the phases of $\varepsilon_{\alpha\beta}$ have been lost, since
there is only one parameter $\cal J$ governing the CP-violating
effects. We stress that, in neglecting the NSIs at sources and
detectors, flavor and mass eigenstates of neutrinos can always be
correlated by using a unitary transformation, and hence, we can use
one effective rephasing invariant to describe the CP-violating
effects in neutrino oscillations. For instance, if we ignore the
source and detector effects in Eq.~(33) of Ref.~\cite{Kopp:2007ne},
the remaining CP-odd terms can be combined together with respect to
a common oscillating factor, which is consistent with our compact
formulas \eqref{eq:PJ1} and \eqref{eq:PJ2}.

Although our exact analytical results are very elegant, they do not
show how new physics affects mixing parameters in a transparent way.
From a phenomenological point of view, analytically approximate
mappings are very useful, since they can reveal the underlying
correlations between effective mixing parameters and NSI effects,
and in particular, show which of them are mostly relevant for a
given process. In the following section, we will perform a detailed
analysis of approximate mappings based on series expansions in small
mixing parameters and NSI corrections. This method is indeed similar
to the analysis of series expansion formulas for neutrino
oscillation probabilities \cite{Akhmedov:2004ny}.

\section{Series expansions of parameter mappings}
\label{sec:expansions}

In this section, we proceed to present the series expansion formulas
of parameter mappings in $\eta$, $s_{13}$, and the NSI parameters
$\varepsilon_{\alpha\beta}$. For convenience, we adopt the standard
parametrization and thus the vacuum leptonic mixing matrix $V$ can
be parametrized by using three mixing angles and one CP violating
phase as
\begin{eqnarray}\label{eq:para}
V & = & O_{23} V_\delta O_{13} V^\dagger_\delta O_{12} \nonumber
\\
&=&\left(
\begin{matrix}c_{12} c_{13} & s_{12} c_{13} & s_{13}
e^{-{\rm i}\delta} \cr -s_{12} c_{23}-c_{12} s_{23} s_{13}
 e^{{\rm i} \delta} & c_{12} c_{23}-s_{12} s_{23} s_{13}
 e^{{\rm i} \delta} & s_{23} c_{13} \cr
 s_{12} s_{23}-c_{12} c_{23} s_{13}
 e^{{\rm i} \delta} & -c_{12} s_{23}-s_{12} c_{23} s_{13}
 e^{{\rm i} \delta} & c_{23} c_{13}\end{matrix}
\right) \ ,
\end{eqnarray}
where $V_\delta = {\rm diag}(1,1,e^{{\rm i}\delta})$, and $O_{ij}$
is the orthogonal rotation matrix in the $(i,j)$-plane with $c_{ij}
\equiv \cos \theta_{ij}$ and $s_{ij} \equiv \sin \theta_{ij}$ (for
$ij=12$, $13$, $23$). A global analysis of current experimental data
yields $0.25 < \sin^2\theta_{12} < 0.37$, $0.36 < \sin^2\theta_{23}
< 0.67$, and $\sin^2\theta_{13} < 0.056$ at the $3\sigma$ confidence
level, but the CP-violating phase $\delta$ is entirely unrestricted
\cite{Schwetz:2008er}. The best-fit values of neutrino mass-squared
differences are $\Delta_{21} = 7.65 \times 10^{-5}~{\rm eV}^2$ and
$|\Delta_{31}| = 2.4 \times 10^{-3}~{\rm eV}^2$, which indicate that
the hierarchy parameter we defined in Sec.~\ref{sec:intro} is given
by $\eta \equiv \Delta_{21} / \Delta_{31} \simeq \pm 0.032$. Present
experimental bounds on the NSI parameters $\varepsilon_{\alpha
\beta}$ show that $\varepsilon_{\mu \alpha }$ (or
$\varepsilon_{\alpha \mu}$) are strongly constrained to
$|\varepsilon_{e\mu}| \lesssim 3.8 \times 10^{-4}$ and $-0.05 <
\varepsilon_{\mu\mu}< 0.08$ at $90~\%$ confidence level
\cite{Ribeiro:2007ud}. This is the reason why some of the previous
works neglect contributions of $\varepsilon_{\mu \alpha}$
\cite{Blennow:2008eb}. In the following, we {will} only focus on
$\varepsilon_{e\tau}$, $\varepsilon_{\mu\tau}$,
$\varepsilon_{\mu\mu}$, and $\varepsilon_{\tau\tau}$ contributions,
respectively.

Using a similar notation, one may also define the effective mixing
angles
$\tilde{\theta}_{12},\tilde{\theta}_{13},\tilde{\theta}_{23}$, and
CP violating phase $\tilde{\delta}_{}$ in matter. Then, we can
parameterize $\tilde{V}$ in analogy to Eq.~\eqref{eq:para}. It is
straightforward to extract the sines of the mixing angles from
Eq.~\eqref{eq:para} using
\begin{eqnarray}\label{eq:angle}
~~~s_{13} =  \left| V_{e3} \right| \; , ~~~~ s_{12} = \left| V_{e2}
\right|/\sqrt{1-\left| V_{e3} \right|^2} \; , ~~~~ s_{23} = \left|
V_{\mu 3} \right|/\sqrt{1-\left| V_{e3} \right|^2} \ .
\end{eqnarray}
The effective mixing angles $\tilde \theta_{ij}$ are obtainable once
the moduli of $\tilde V_{\alpha\beta}$ are computed using
Eq.~\eqref{eq:solution}, and it is not difficult to check that in
the limit of vanishing matter effects the effective mixing angles
are equal to the vacuum ones. Analytical relations between
$\tilde\theta$ and $\theta$ could be very useful and they rely on
the perturbation theory that we have employed.

For our purposes, we first factor out the rotation matrix $O_{23}$
\begin{eqnarray}\label{eq:Hfactor}
\tilde H & = & \frac{\Delta_{31}}{2E}  O_{23}  V_\delta \cdot M
\cdot V^\dagger_\delta O^T_{23}  \nonumber \\ & = &
\frac{\Delta_{31}}{2E} O_{23} V_\delta \cdot \left[ \hat V \cdot
{\rm diag} \left( \lambda_1 , \lambda_2 ,  \lambda_3\right) \cdot
\hat V^\dagger \right] \cdot V^\dagger_\delta O^T_{23} \ ,
\end{eqnarray}
where $M$ is given by
\begin{eqnarray}\label{eq:M}
M =  O_{13} O_{12} \cdot {\rm diag} \left( 0,\eta,1 \right) \cdot
O^T_{12} O^T_{13} + {\rm diag}\left( \hat A,0,0 \right)  +
V^\dagger_\delta O^T_{23} \cdot \varepsilon \cdot O_{23} V_\delta \
,
\end{eqnarray}
and $\hat A \equiv A/\Delta_{31}$. In deriving
Eq.~\eqref{eq:Hfactor}, the following commutative properties are used
\begin{eqnarray}\label{eq:commute}
V^\dagger_\delta O_{12} & = & O_{12} V^\dagger_\delta \ , \\
V^\dagger_\delta \cdot {\rm diag}\left( \hat A,0,0 \right) & = &
{\rm diag}\left( \hat A,0,0 \right) \cdot V^\dagger_\delta \ .
\end{eqnarray}

The diagonalization of $M$ is performed by using perturbation
theory, i.e., we write $M= M^{(0)} + M^{(1)} + \cdots$, where
$M^{(1)}$ contains all terms of first order in $\eta$, $s_{13}$, and
$\varepsilon_{\alpha\beta}$. One finds
\begin{eqnarray}\label{eq:M0}
M^{(0)} & = &  {\rm diag} \left( \hat A , 0 , 1\right) = {\rm diag}
\left( \lambda^{(0)}_1 ,  \lambda^{(0)}_2 ,  \lambda^{(0)}_3\right)
\ ,
\end{eqnarray}
and
\begin{eqnarray}\label{eq:M1}
M^{(1)} & = & \left(\begin{matrix} \eta s^2_{12} + \hat A
\hat\varepsilon_{ee} & \eta s_{12} c_{12} + \hat A
\hat\varepsilon_{e\mu}  &   s_{13} e^{-{\rm i}\delta} + \hat
A\hat\varepsilon_{e\tau} \cr \sim & \eta c^2_{12} + \hat
A\hat\varepsilon_{\mu\mu} & \hat A\hat \varepsilon_{\mu\tau} \cr
\sim & \sim & \hat A\hat\varepsilon_{\tau\tau}
\end{matrix}\right) \ ,
\end{eqnarray}
with `$\sim$' denoting the conjugate elements and
$\hat\varepsilon_{\alpha\beta} = (V^\dagger_\delta O^T_{23}  \cdot
\varepsilon \cdot  O_{23} V_\delta )_{\alpha\beta}$. Since $M^{(0)}$
is diagonal at zeroth order, we have $\hat V^{(0)}=I$. Then, the
first order corrections are given by
\begin{eqnarray}\label{eq:lambda1}
\lambda^{(1)}_{i} & = &  M^{(1)}_{ii} \ ,
\end{eqnarray}
and
\begin{eqnarray}\label{eq:V1}
\hat{V}^{(1)}_i & = & \sum_{j\neq i} \frac{ M^{(1)}_{ji}
}{\lambda^{(0)}_i - \lambda^{(0)}_j} \hat{V}^{(0)}_j \ .
\end{eqnarray}
{Thus,} the effective masses and mixing matrix are given by $\tilde
m^2_i \simeq \Delta_{31} ( \lambda^{(0)}_i + \lambda^{(1)}_i)$ and
$\tilde V \simeq O_{23} V_\delta (\hat V^{(0)} +\hat V^{(1)} )$,
respectively. Finally, inserting Eq.~\eqref{eq:M1} into
Eqs.~\eqref{eq:lambda1} and \eqref{eq:V1}, we arrive at mappings for
the effective mass squares
\begin{eqnarray}\label{eq:approxm}
\tilde m^2_1 & \simeq & \Delta_{31} \left(\hat A + \eta s^2_{12} +
\hat A\varepsilon_{ee}\right) \ , \\
\tilde m^2_2 & \simeq & \Delta_{31} \left[ \eta c^2_{12} - \hat A
s^2_{23} \left(\varepsilon_{\mu\mu}-\varepsilon_{\tau\tau}\right) -
\hat A s_{23} c_{23} \left(\varepsilon_{\mu\tau} +
\varepsilon^*_{\mu\tau}\right) + \hat A \varepsilon_{\mu\mu} \right]
\ , \\ \tilde m^2_3 & \simeq & \Delta_{31} \left[ 1 + \hat A
\varepsilon_{\tau\tau} + \hat A s^2_{23}
\left(\varepsilon_{\mu\mu}-\varepsilon_{\tau\tau}\right) + \hat A
s_{23} c_{23} \left(\varepsilon_{\mu\tau} +
\varepsilon^*_{\mu\tau}\right)\right] \ ,
\end{eqnarray}
the effective mixing matrix elements,
\begin{eqnarray}
\label{eq:approxV1} \tilde V_{e2} & \simeq & \frac{\eta s_{12}
c_{12}}{\hat A} + c_{23}\varepsilon_{e\mu} - s_{23}
\varepsilon_{e\tau}  \ , \label{eq:approxV2} \\ \tilde V_{e3} &
\simeq &  \frac{s_{13}e^{-{\rm i}\delta}}{1-\hat A} + \frac{\hat A
(s_{23}\varepsilon_{e\mu}+ c_{23}\varepsilon_{e\tau})  }{1-\hat A} \
,  \\ \tilde V_{\mu 2} & \simeq & c_{23} + s^2_{23} c_{23} \hat A
\left( \varepsilon_{\tau\tau} - \varepsilon_{\mu\mu} \right) +
s_{23} \hat A \left( s_{23} \varepsilon_{\mu\tau} -c^2_{23}
\varepsilon^*_{\mu\tau} \right) \ ,
\\ \tilde V_{\mu 3} & \simeq &
s_{23} + \hat A  \left[ s_{23} \left( \varepsilon_{\mu\mu} -
\varepsilon_{\tau\tau} \right)  + c_{23} \varepsilon_{\mu\tau} -
s^2_{23} c_{23}  \left( \varepsilon_{\mu\tau} +
\varepsilon^*_{\mu\tau} \right) +  s^3_{23} \left(
\varepsilon_{\tau\tau} - \varepsilon_{\mu\mu} \right) \right] \ ,
\label{eq:approxV3}
\end{eqnarray}
and the effective Jarlskog parameter
\begin{eqnarray}\label{eq:Jeff}
\tilde J & = & \frac{s_{13} s_{23} \left(\eta s_\delta c_{12} c_{23}
s_{12} + \hat A s_{\delta +\phi _{{e\mu }}} c^2_{23} |\varepsilon
_{{e\mu }}| - \hat A s_{\delta +\phi _{{e\tau }}} c_{23}
s_{23} |\varepsilon _{{e \tau }}| \right)}{(\hat A-1) \hat A} \nonumber \\
&+& \frac{s_{23} \left(s_{\phi _{{e\mu }}-\phi _{{e\tau }}} c_{23}
|\varepsilon _{{e\mu }}| |\varepsilon _{{e\tau }}| \hat A^2-\eta
s_{\phi _{{e\mu }}} c_{12} c_{23} s_{12} s_{23} |\varepsilon _{{e\mu
}} |\hat A - \eta s_{\phi _{{e\tau }}} c_{12} s_{12} c^2_{23}
|\varepsilon _{{e\tau }}| \hat A\right)}{(\hat A-1) \hat A} \ ,
\end{eqnarray}
where the $\phi_{\alpha\beta}$'s are the phases associated with the
complex NSI parameters $\varepsilon_{\alpha\beta}$ and {the
$s_{\phi_{\alpha\beta}}$'s are} the corresponding sine functions.
The above mappings can be transferred into mappings between mixing
angles straightforwardly, i.e., we have approximately $\tilde s_{13}
\simeq |\tilde V_{e3}|$,  $\tilde s_{12} \simeq |\tilde V_{e2}|$,
and $\tilde s_{23} \simeq |\tilde V_{\mu3}|$.
Equations~\eqref{eq:approxm}-\eqref{eq:Jeff} are our main results
for the approximate analytical mappings.

Some discussions are in order:
\begin{itemize}

\item In the limit $\varepsilon_{\alpha \beta} \to 0$, it is
  interesting to observe that our results coincide with the mapping
  results in Ref.~\cite{Freund:2001pn} when evaluated at the same
  order of perturbation theory. In fact, expanding Eqs.~(27a)-(27c) of
  Ref.~\cite{Freund:2001pn} in $\eta$ and $s_{13}$ (which means that
  the $\hat C$ parameter appearing there equals $1-\hat A$), and retaining
  terms up to first order in these small parameters, we can easily
  check that they reduce to
\begin{eqnarray}\label{limit}
 \tilde s_{13} &=&  \frac{s_{13}  }{1-\hat A}  \ , \\
\tilde s_{23} &=&  s_{23}  \ ,  \\
\tilde s_{12} &=& \frac{\eta}{\hat A} c_{12} \,s_{12} \ ,
\end{eqnarray}
which are in agreement with our
Eqs.~\eqref{eq:approxV1}-\eqref{eq:approxV3} evaluated at
$\varepsilon_{\alpha \beta}=0$. In addition, from Eq.~\eqref{limit},
it is worth noticing that the mixing parameter $s_{13}$ is strongly
modified by matter effects when $0 < \hat A \ne 1$, otherwise the
resonance $\hat A = 1$ is at work and the mapping procedure adopted
in this paper is not valid.

\item As shown in Eq.~\eqref{eq:commute}, the orthogonal matrix
  $O_{23}$ commutes with the standard matter potential. Hence, in
  the case of vanishing $\theta_{13}$ and NSIs, Eq.~\eqref{eq:Hm} can
  be rewritten as
\begin{eqnarray}\label{Hs130}
\tilde{H} & = &  \frac{\Delta_{31}}{2E}O_{23}\left[ O_{12} \cdot
{\rm diag} \left(0,\eta,1 \right) \cdot O^T_{12} + {\rm
diag} \left(\hat A,0,0\right) \right] O^T_{23}  \ ,
\end{eqnarray}
where the CP violating phase $\delta$ loses its meaning and does not
appear. An evident conclusion deduced from Eq.~\eqref{Hs130} is
that, if $\theta_{13}=0$, the standard matter effects only
contribute to the mixing angle ${\theta}_{12}$. Both $\theta_{13}$
and $\theta_{23}$ as well as $\Delta_{31}$ will keep their vacuum
values in matter. However, when the NSIs are taken into account, the
situation will be quite different. A non-zero $\tilde{\theta}_{13}$
will emerge in general, $\tilde{\theta}_{23}$ will deviate from its
maximal value $\pi/4$, and a non-trivial CP violating phase $\tilde
\delta $ may also exist.

\item  As already noticed above, the mapping for
$\theta_{12}$ shows an unphysical divergence for $\hat A \to 0$ or
$\hat A \to 1$, and the vacuum results cannot be reproduced, a
well-known consequence of the perturbative approach adopted in the
mapping procedure. Thus, degenerate perturbation theory should be
elaborated on around these two singularities.

\item Except from $\varepsilon_{e\mu}$ and $\varepsilon_{e\tau}$, it
can also be very clearly seen that contributions to $\tilde s_{13}$
from all the other NSI parameters are all suppressed. Since the
present experimental bound on $\varepsilon_{e\mu}$ is rather
stringent \cite{Davidson:2003ha}, we conclude that
$\varepsilon_{e\tau}$ is the most significant NSI parameter to be
taken into account for $\tilde \theta_{13}$. As for $\tilde
\theta_{23}$, NSI corrections  are  relatively mild unless very
high-energy regions are considered.

\item The matrix elements $|\tilde V_{\mu 3}|$, $|\tilde V_{\tau 3}|$,
$|\tilde V_{\mu 1}|$, and $|\tilde V_{\tau 1}|$
are not modified by $\varepsilon_{e\tau}$.
\end{itemize}

\section{Applications}
\label{sec:applications}

We now proceed to numerically illustrate (using normal mass hierarchy,
i.e., $\eta \simeq + 0.032$) the NSI corrections to the leptonic
mixing parameters and the neutrino oscillations probabilities based on
our model independent results obtained in Secs.~\ref{sec:sum_rules}
and \ref{sec:expansions}. We first consider the exact results for
mixing matrix elements, and combinations of them, as obtained applying
Eq.~\eqref{eq:solution}.  This is an important point, since both
mixing angles and the Jarlskog parameter depend on the modified
behavior of the effective matrix elements [see Eqs.~\eqref{eq:Jm} and
\eqref{eq:angle}] due to matter effects and non-standard physics. We
then focus on the importance of the NSI effects on $\theta_{13}$ and
$\theta_{23}$ in correcting their tri-bimaximal mixing values (i.e.,
$\theta_{13}=0$ and $\theta_{23}=\pi/4$)
\cite{Harrison:2002er,Xing:2002sw}.  Finally, in order to show the
goodness of our approximate results for the effective mixing angles,
we compare the exact $\nu_e \to \nu_\mu$, $\nu_e \to \nu_\tau$, and
$\nu_\mu \to \nu_\tau$ oscillation probabilities obtained including
the NSI effects with those derived using
Eqs.~\eqref{eq:approxV1}-\eqref{eq:approxV3} for the effective mixing
angles.

\subsection{NSI corrections to the leptonic mixing matrix}

\begin{figure}[t!]
\begin{center}
\vspace{-4.6cm}
\epsfig{file=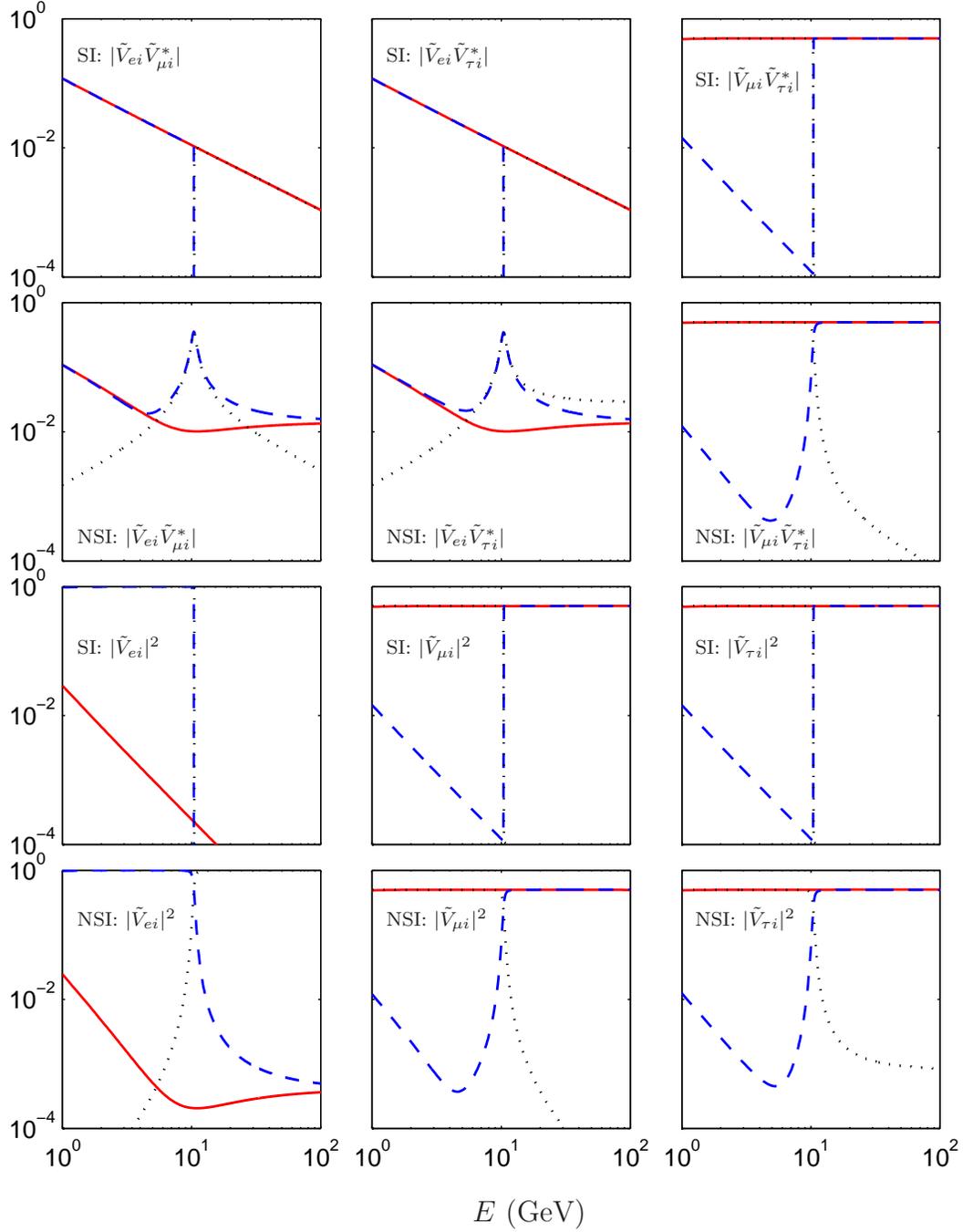,bbllx=8.cm,bblly=27cm,bburx=12.cm,bbury=31cm,%
width=4.1cm,height=4.1cm,angle=0,clip=0}\vspace{18.2cm} \caption{\it
Illustrative plots for the effective matrix elements $|\tilde
V_{\alpha i}\tilde V^{*}_{\beta i}|$ (first and second rows) and
$|\tilde V_{\alpha i}|^2$ (third and fourth rows) as a function of the
neutrino energy $E$. Here the solid, dashed, and dotted {curves}
correspond to $i=1,2,3$, respectively. The first and third rows show
the results without including NSIs (labeled SI), while the second and
fourth rows are those including NSIs. We use the representative value
${\rm Re}(\varepsilon_{e\tau})={\rm Im}(\varepsilon_{e\tau})=0.02$,
with all other $\varepsilon_{\alpha\beta}$ being zero.}\vspace{-0cm}
\label{fig:fig1}
\end{center}
\end{figure}

The effective leptonic mixing matrix can be reconstructed directly
from Eq.~\eqref{eq:solution}. For $\alpha=\beta$, we obtain the
expressions of the matrix elements $|\tilde V_{\alpha i}|$, whereas
for $\alpha \neq \beta$, we obtain the sides of unitarity triangles
$\tilde V_{\alpha i} \tilde V^{*}_{\beta i}$. The numerical results
are presented in Fig.~\ref{fig:fig1}. In our numerical calculations,
we take the central values of the neutrino mass-squared differences
($\Delta_{21} = 7.65 \times 10^{-5} ~ {\rm eV}^{2}$, $\Delta_{31} =
2.4 \times 10^{-3} ~ {\rm eV}^{2}$) and {the leptonic} mixing angles
($\theta_{12} = 33.5^\circ$, $\theta_{23} = 45^\circ$,
$\theta_{13}=0$) obtained in a global analysis of the presently
available neutrino oscillation data \cite{Schwetz:2008er}. Here,
just as an example, we choose $\varepsilon_{e\tau}$ as the only
non-vanishing NSI parameter. For comparison, we also show the
results without including NSIs.
%%%%%%%%%%%%%%%%%%%% Fig. 1 %%%%%%%%%%%%%%%%
%%%%%%%%%%%%%%%%%%%%%%%%%%%%%%%%%%%%%%%%%%%
In particular, for higher neutrino beam energies, the NSI corrections
are remarkable.

First, we can observe that the energy dependence of the matrix
elements can be easily read off from the approximate relations in
Eqs.~\eqref{eq:approxV1}-\eqref{eq:approxV3}, since the matter
parameter $A \sim E$.\footnote{The matrix elements {that are} not
quoted in Eqs.~\eqref{eq:approxV1}-\eqref{eq:approxV3} can be
obtained using unitarity relations.} Thus, for example, the fact
that $|\tilde V_{e 2}|$ is predicted to decrease with increasing
neutrino energy is confirmed in the first panel of the fourth row in
Fig.~\ref{fig:fig1}. Moreover, the singularity for matrix elements
around $E\sim 10~{\rm GeV}$ (see panels in the first and third rows)
clearly corresponds to the resonance at $ \hat A\sim1$, which can be
understood investigating the perturbative results in
Eqs.~\eqref{eq:approxV1}-\eqref{eq:approxV3}. Note that, for
anti-neutrinos, since the sign in front of $\hat A$ is negative,
such a singularity should not appear. In addition, the relation
$|\tilde V_{e i} \tilde V^{*}_{\mu i}| \simeq |\tilde V_{e i}\tilde
V^{*}_{\tau i}|$ holds quite well, which is an obvious consequence
of the $\mu-\tau$ symmetry in the genuine neutrino mass matrix. This
can be seen comparing the first and second panels of the second row
for any value of the index $i$. In the same panels, at energies
$E\sim4 ~ {\rm GeV}$, $\tilde V_{e i} \tilde V^{*}_{\mu i}$ and
$\tilde V_{e i} \tilde V^{*}_{\tau i}$ ($i=1,2,3$) are comparable to
each other, and thus, the unitarity triangle built with these sides
takes a nearly equilateral form with three nearly degenerate inner
angles.  Such an equilateral form is destroyed when increasing the
energy. Similarly, $|\tilde V_{\mu 1} \tilde V^{*}_{\tau 1}|$ is
rather stable against matter corrections and NSI effects, which also
reflects the stabilization of $\tilde \theta_{23}$. As for the
matrix elements, $|\tilde V_{\mu 1}|$ and $|\tilde V_{\tau 1}|$ are
not sensitive to $\varepsilon_{e\tau}$, which is also in agreement
with our approximate mappings.

\subsection{NSI corrections to the mixing angles}

A crucial goal of future neutrino facilities is to measure the
smallest leptonic mixing angle $\theta_{13}$ in order to extract
information on leptonic CP violation. However, it has been pointed
out that NSIs may play a very important role for mimicking effects
on $\theta_{13}$ and leptonic CP violation, especially in the case
of a small $\theta_{13}$ \cite{Huber:2001de}. It is then quite
important to analyze in detail these  effects in order to be able to
disentangle
 genuine $\theta_{13}$ effects from new
physics-induced ones at future neutrino facilities.

On the other hand, the question of whether the leptonic mixing angle
$\tilde \theta_{23}$ is exactly maximal or not is quite relevant,
especially from the model builders' point of view: in fact, many
models presented in the literature predict $\theta_{23}$ being
(almost) maximal and the understanding of the flavor problem
strongly relies on the knowledge on the value of $\theta_{23}$ to be
as accurate as possible. Thus, it is very important to investigate
the possible NSI corrections to $\theta_{13}$ and the maximal mixing
pattern in the $\mu-\tau$ sector.

According to Eqs.~\eqref{eq:approxV1}-\eqref{eq:approxV3}, the most
relevant NSI parameter for $\tilde \theta_{13}$ is
$\varepsilon_{e\tau}$ (since the upper bound on $\varepsilon_{e\mu}$
is rather stringent), whereas for $\tilde \theta_{23}$
$\varepsilon_{\mu\tau}$, $\varepsilon_{\mu\mu}$ and
$\varepsilon_{\tau\tau}$ contribute. Notice that, in the latter case
for maximal mixing ($\theta_{23}=45^\circ$), a typical feature is
that the vacuum Hamiltonian takes on a $\mu-\tau$ symmetric form,
namely, $H$ is invariant under the exchange of $\mu$ and $\tau$
indices. Hence, if NSIs possess a similar $\mu-\tau$ symmetric form
(i.e., $\varepsilon_{e\mu} = \varepsilon_{e\tau} $ and
$\varepsilon_{\mu\mu} = \varepsilon_{\tau\tau}$), the $\mu-\tau$
symmetry exists in the effective Hamiltonian $\tilde H$, and the
effective mixing angle $\tilde \theta_{23} $ will not be affected by
matter effects. As a consequence, $\varepsilon_{\mu\tau}$ itself
does not contribute to $\tilde \theta_{23}$ if all the other NSI
parameters are zero.

%%%%%%%%%%%%%%%%%%%% Fig. 2 %%%%%%%%%%%%%%%%
\begin{figure}[t]
\begin{center}
\vspace{-1.5cm}
\epsfig{file=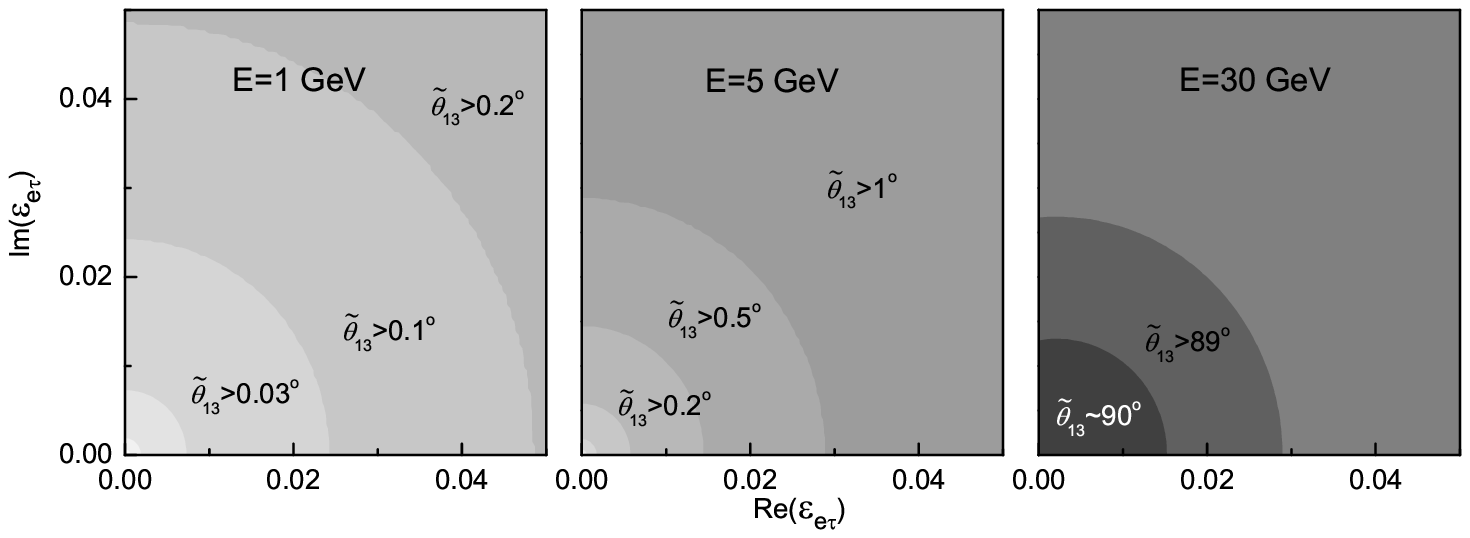,bbllx=5cm,bblly=3cm,bburx=9cm,bbury=7cm,%
width=4cm,height=4cm,angle=0,clip=0}
\epsfig{file=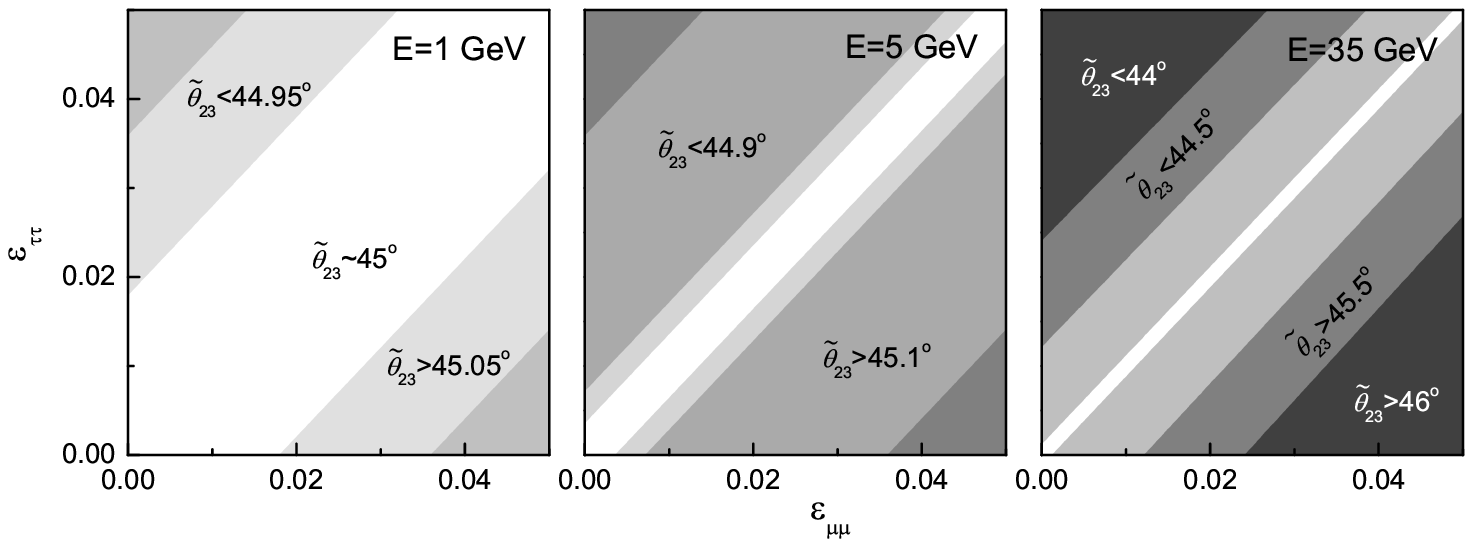,bbllx=9.2cm,bblly=9cm,bburx=13.2cm,bbury=13cm,%
width=4cm,height=4cm,angle=0,clip=0} \vspace{9cm} \caption{\it
Dependence of the mixing angles $\tilde \theta_{13}$ and $\tilde
\theta_{23}$ on the NSI parameters for three representative values
of the neutrino energy: $E=1,5$, and $30~(35)$ GeV, which roughly
correspond to the $\nu_e$ and $\nu_\mu$ mean energies at a 50 GeV
neutrino factory. The vacuum value of $\theta_{13}$ is fixed to be
zero, whereas we assume maximal mixing for $\theta_{23}$. In each
plot, the darker the region, the larger the deviation of $\tilde
\theta_{13}$ and $\tilde \theta_{23}$ from their vacuum values. Only
the labeled NSI parameters are non-vanishing in each plot.}
\label{fig:fig2}
\end{center}
\end{figure}
%%%%%%%%%%%%%%%%%%%%%%%%%%%%%%%%%%%%%%%%%%%

%%%%%%%%%%%%%%%%%%%% Fig.~x %%%%%%%%%%%%%%%%
\begin{figure}[t]
\begin{center}
\vspace{-4.cm}
\epsfig{file=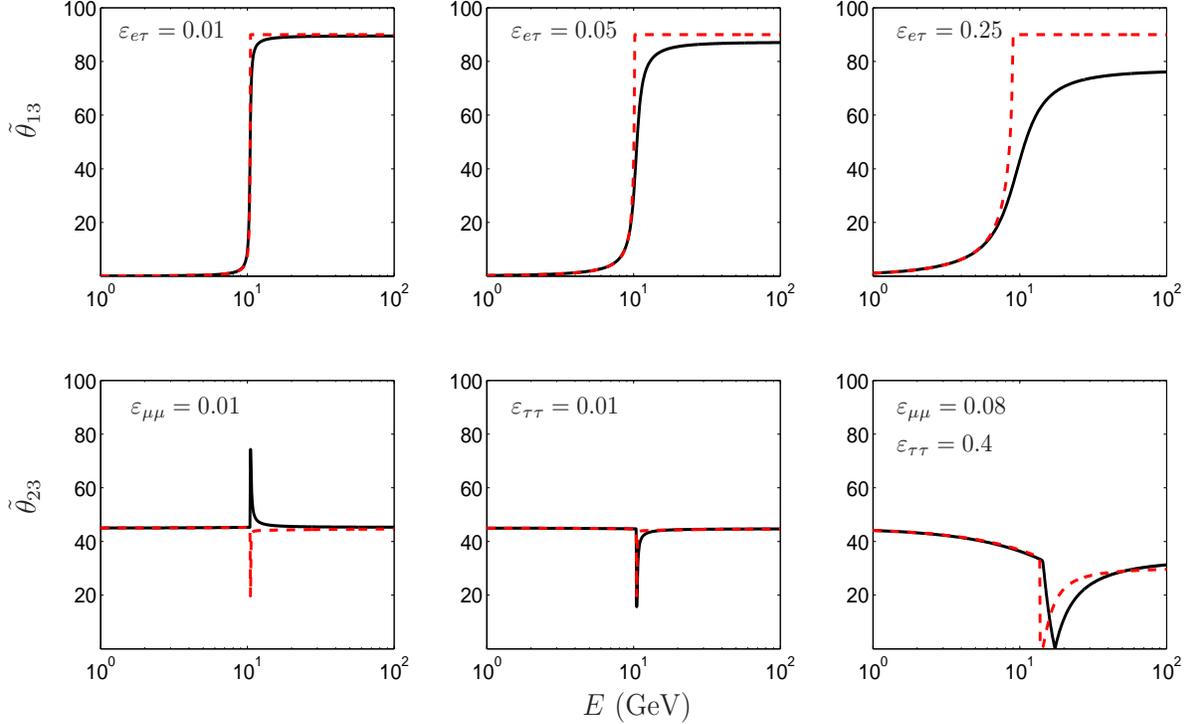,bbllx=7.5cm,bblly=23cm,bburx=12.5cm,bbury=27cm,%
width=4.5cm,height=3.8cm,angle=0,clip=0} \vspace{10cm}
\caption{\it Neutrino energy dependence of the  effective mixing angles ($\tilde
\theta_{13}$ and $\tilde\theta_{23}$).
Solid curves correspond to exact numerical results, whereas the dashed ones are
computed using our approximate mappings. The non-vanishing NSI
parameters have been labeled in each plot.} \label{fig:figx}
\vspace{0.5 cm}
\end{center}
\end{figure}
%%%%%%%%%%%%%%%%%%%%%%%%%%%%%%%%%%%%%%%%%%%
%%%%%%%%%%%%%%%%%%%%%%%%%%%%%%%%%%%%%%%%%%%

In the upper plots of Fig.~\ref{fig:fig2}, we show the non-vanishing
$\tilde{\theta}_{13}$ generated by the NSIs [computed using our
exact formula given in Eq.~\eqref{eq:solution}]. One can observe
that $\tilde{\theta}_{13}$ is quite sensitive to
$\varepsilon_{e\tau}$. In the case of $E=30~{\rm GeV}$,
$\tilde{\theta}_{13}$ may acquire a very sizable value close to
$90^\circ$. This is due to the reordering of the eigenvalues
$\tilde{m}_1$ and $\tilde m_3$ when $E \gtrsim 10 ~{\rm GeV}$. If we
keep the order of eigenvalues in the form of ${\rm diag} (\tilde
m^2_1,\tilde m^2_2,\tilde m^2_3)$, a shift of $\pi/2$ has to be
added to $\tilde \theta_{13}$. In the lower plots of
Fig.~\ref{fig:fig2}, we show the corrections from
$\varepsilon_{\mu\mu}$ and $\varepsilon_{\tau\tau}$ to
$\theta_{23}$, assuming the vacuum value $\theta_{23}=\pi/4$. Our
numerical results indicate that there are no sizable corrections to
the mixing angle $\theta_{23}$, and even in the high-energy region,
$\tilde{\theta}_{23}$ should not deviate from its maximal value by
more than a few degrees. We also find that the
$\varepsilon_{\mu\mu}-\varepsilon_{\tau\tau}$ contributions are
symmetric with respect to the $ \varepsilon_{\mu \mu} =
\varepsilon_{\tau \tau} $ axis up to a minus sign, since an addition
to one of the parameters could as well be made to the other one.

It is interesting to observe that the main features of the previous
{\it exact} results can also be captured from our approximate mappings
in Eqs.~\eqref{eq:approxV1}-\eqref{eq:approxV3}. To illustrate this
point, we show the dependence of $\tilde \theta_{13}$ and $\tilde
\theta_{23}$ on the neutrino energy in Fig.~\ref{fig:figx}, for
different values of the relevant NSI parameters, according to our
foregoing discussions. Solid curves correspond to exact results
obtained using Eq.~\eqref{eq:solution}, whereas dashed ones represent
our perturbative mappings. In the first row, we can appreciate how the
dependence of $\tilde \theta_{13}$ on $\varepsilon_{e\tau}$ is well
described by our perturbative result in Eq.~\eqref{eq:approxV2},
unless $\varepsilon_{e\tau}$ assumes a very large value, close to its
upper bound \cite{Ribeiro:2007ud}. In addition, notice that the
increase of $\tilde \theta_{13}$ corresponds to reordering the
eigenvalues for energies around 10 GeV. In the second row, we analyze
the behavior of $\tilde \theta_{23}$, for different values of the
relevant parameters $\varepsilon_{\mu\mu}$ and
$\varepsilon_{\tau\tau}$. In particular, in the first and second
panels, we choose only one of them being different from zero (and
equal to $0.01$), whereas in the last one, we allow both of them to
assume larger values ($\varepsilon_{\mu\mu}=0.08$ and
$\varepsilon_{\tau\tau}=0.4$). The agreement between our calculation
and the exact evaluation of $\tilde \theta_{23}$ is quite good, also
in predicting the location of the resonance.

Finally, we comment on the fact that $\tilde \theta_{12}$ is
dramatically suppressed by matter effects, as shown in
Eq.~\eqref{eq:approxV2}. However, since long-baseline neutrino
oscillation experiments are not very sensitive to this angle, we
will not perform a detailed analysis here. The conclusions made
above about the dependence of the effective angles on the NSI
parameters apply as well to the case of anti-neutrinos in matter and
we will not perform such a redundant analysis here.

\subsection{NSI corrections to the neutrino oscillation probabilities}

Until now we have described the relevant features of new physics
effects on the matrix elements of the leptonic mixing matrix. In
what follows, we study the dependence of the transition
probabilities on the NSI parameters, for different choices of
neutrino energies and baselines. In particular, we focus on the {\it
golden channel} $\nu_e\rightarrow \nu_\mu$ \cite{Cervera:2000kp} and
the CP asymmetries derived from it, the {\it silver  channel}
$\nu_e\rightarrow \nu_\tau$ \cite{Donini:2002rm,Autiero:2003fu}, and
the so-called {\it discovery channel} $\nu_\mu\rightarrow \nu_\tau$,
which is thought to be the best channel for searching for new
physics \cite{Donini:2008wz}. We also show how the relevant features
of the transition probabilities are well reproduced computing them
by inserting Eqs.~\eqref{eq:approxV1}-\eqref{eq:approxV3} into
Eq.~\eqref{eq:Pm}.

\begin{figure}[t]
\begin{center}\vspace{-4.cm}
\epsfig{file=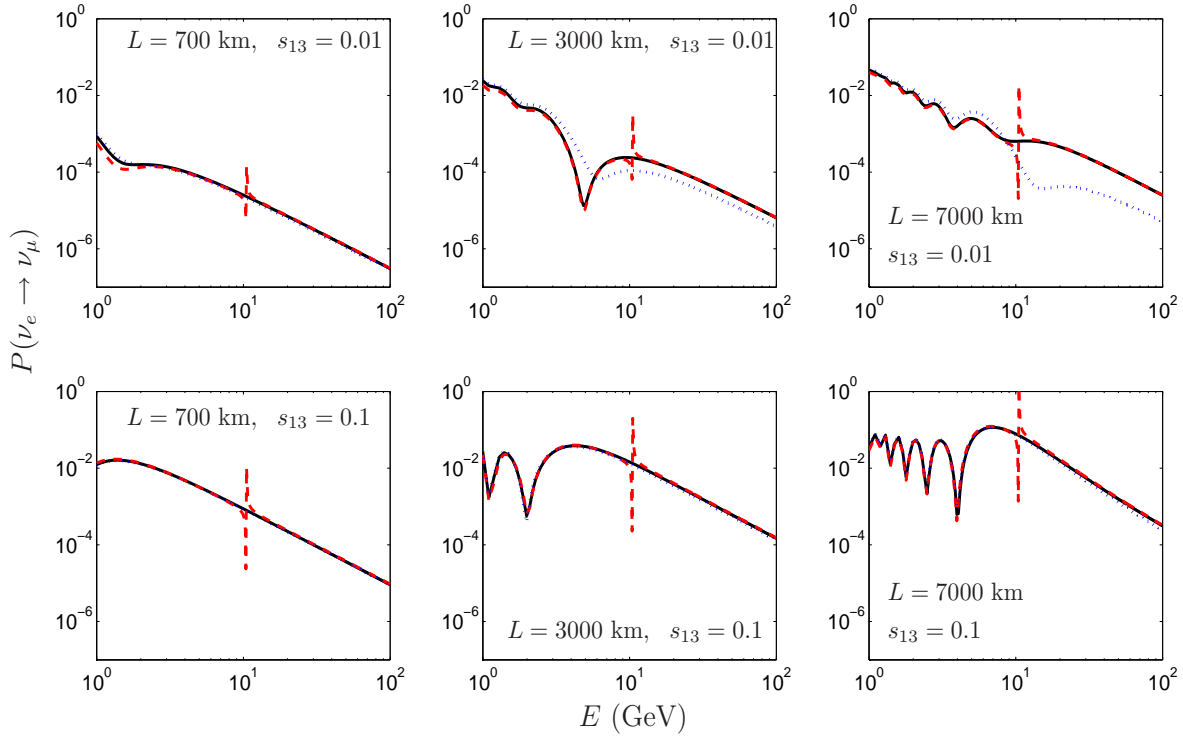,bbllx=7.5cm,bblly=23cm,bburx=12.5cm,bbury=27cm,%
width=4.5cm,height=3.8cm,angle=0,clip=0} \vspace{10cm} \caption{\it
Neutrino oscillation probabilities for the $\nu_e\rightarrow
\nu_\mu$ channel as a function of the neutrino energy $E$. The
baseline lengths and values of $s_{13}$ have been labeled
in each plot. Here, we set $\delta = \pi/2$ and only
$\varepsilon_{e\tau}=0.01$ is allowed to be non-vanishing. The solid
curves denote the exact numerical results. The dashed curves
correspond to results derived from our approximate mappings, and for
comparison, the dotted curves are shown to illustrate probabilities
without including NSIs.}\label{fig:fig3}
\end{center}
\end{figure}

In Fig.~\ref{fig:fig3}, we show the transition probability $P(\nu_e
\rightarrow \nu_\mu)$ as a function of the neutrino energy for three
different baseline setups: $L=700~{\rm km}$ (around the scope of
MINOS \cite{Michael:2006rx} and OPERA \cite{Guler:2000bd}),
$L=3000~{\rm km}$, and $L=7000~{\rm km}$ (for the two detector setup
of a neutrino factory).
%%%%%%%%%%%%%%%%%%%% Fig.~3 %%%%%%%%%%%%%%%%
%%%%%%%%%%%%%%%%%%%%%%%%%%%%%%%%%%%%%%%%%%%
%%%%%%%%%%%%%%%%%%%%%%%%%%%%%%%%%%%%%%%%%%%
The input parameters are the same as those in Fig.~\ref{fig:fig1}.  In
each panel, the solid curves denote the exact numerical results, the
dashed curves correspond to results derived from our approximate
mappings and, to highlight the effects of the NSI parameters, the
dotted curves represent the probability without including NSIs. We can
observe that our approximate mappings given in
Sec.~\ref{sec:expansions} agree with the exact numerical results to an
extremely good precision. Similar to the plots of the mixing
parameters, a singularity exists around $E\sim 10 ~{\rm GeV}$ due to
the limitation of non-degenerate perturbation theory that we have
elaborated. For smaller $\theta_{13}$, the probability is more
sensitive to the NSI effects, and thus, longer baseline lengths are
more favored for the purpose of searching for new physics effects.

\begin{figure}[t]
\begin{center}
\vspace{-4cm}
\epsfig{file=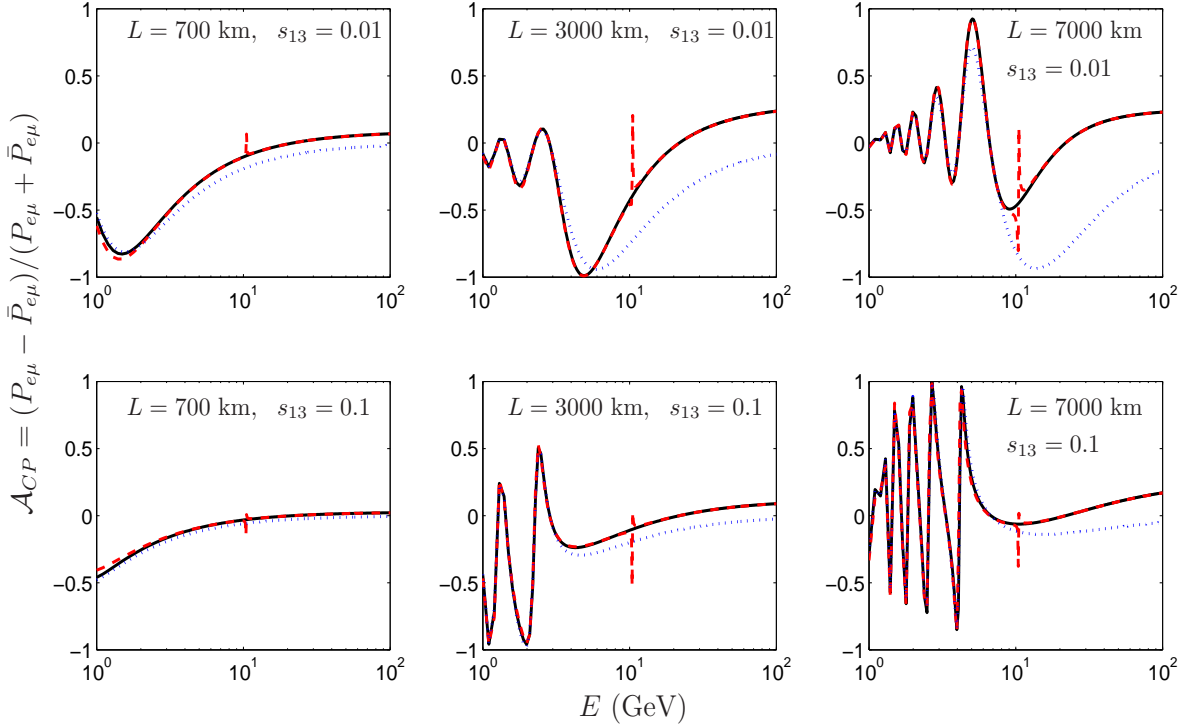,bbllx=7.5cm,bblly=23cm,bburx=12.5cm,bbury=27cm,%
width=4.5cm,height=3.8cm,angle=0,clip=0} \vspace{10cm} \caption{\it
CP asymmetry ${\cal A}_{CP}$ derived from the $\nu_e\rightarrow
\nu_\mu$ channel. The values of the mixing parameters as well as
those of the baselines and neutrino energies are the same as in
Fig.~\ref{fig:fig3}. The solid curves denote the exact numerical
results, the dashed curves correspond to results derived from our
approximate mappings, and the dotted curves show the probabilities
without including NSIs.} \label{fig:fig4} \vspace{0 cm}
\end{center}
\end{figure}

The experimentally measured CP asymmetry in the golden channel,
which is usually defined as
\begin{eqnarray}\label{eq:ACP}
{\cal A}_{CP} = \frac{P(\nu_e \rightarrow \nu_\mu)-P(\overline \nu_e
\rightarrow \overline \nu_\mu)}{P(\nu_e \rightarrow
\nu_\mu)+P(\overline \nu_e \rightarrow \overline \nu_\mu)} \ ,
\end{eqnarray}
is illustrated in Fig.~\ref{fig:fig4} for the same baseline
setup.
%%%%%%%%%%%%%%%%%%%%%%%%%%%%%%%%%%%%%%%%%%%
%%%%%%%%%%%%%%%%%%%% Fig.~4 %%%%%%%%%%%%%%%%
%%%%%%%%%%%%%%%%%%%%%%%%%%%%%%%%%%%%%%%%%%%
%%%%%%%%%%%%%%%%%%%%%%%%%%%%%%%%%%%%%%%%%%%
Again, our approximate mappings are valid in a large range of beam
energies. At higher energies, the CP asymmetries are dramatically
affected by NSIs, i.e., the values of ${\cal A}_{CP}$, which are
calculated without taken into account NSI effects, may go to
divergent directions.

\begin{figure}[t]
\begin{center}
\vspace{-4.5cm}
\epsfig{file=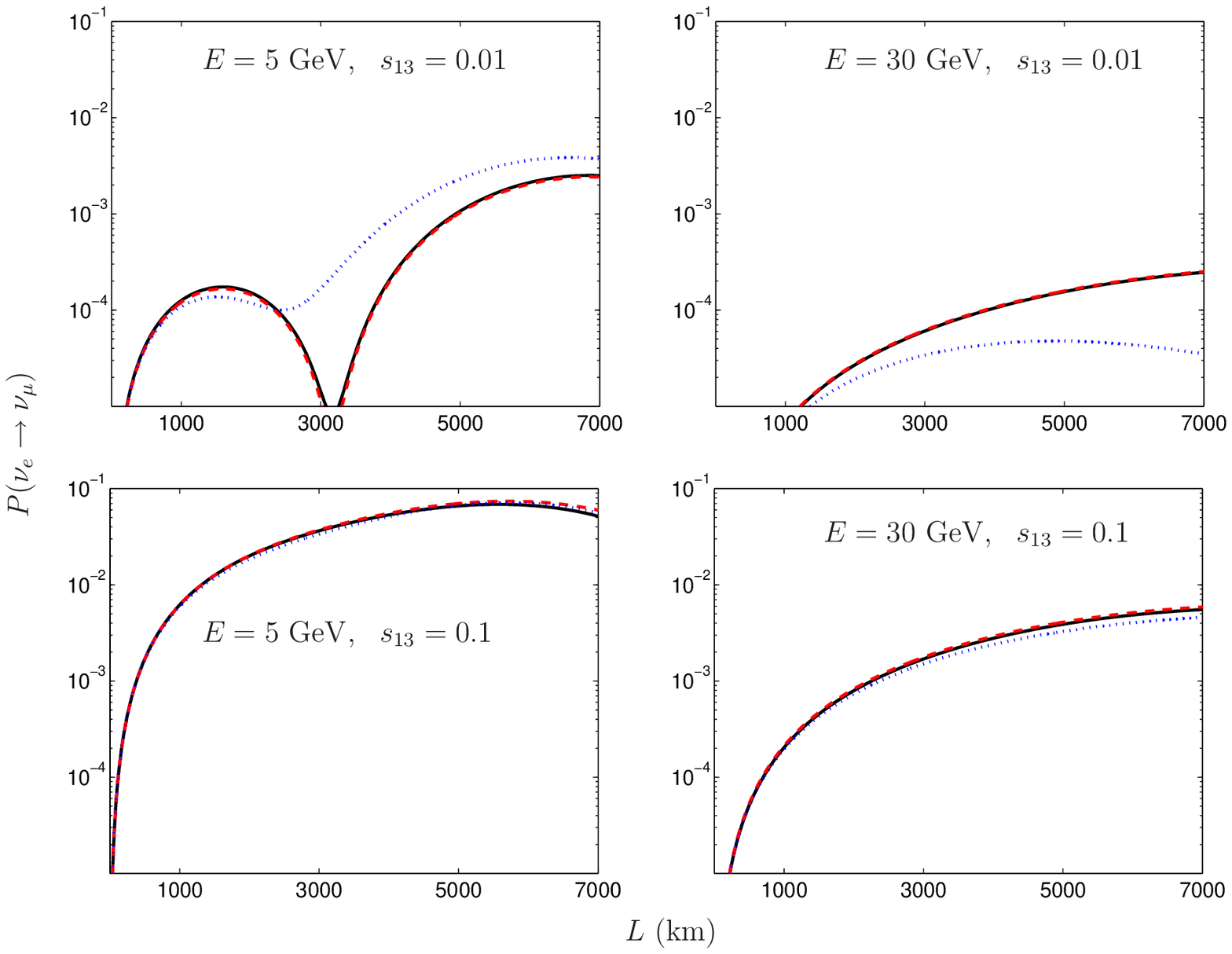,bbllx=9cm,bblly=23cm,bburx=13cm,bbury=27cm,%
width=3.cm,height=3.cm,angle=0,clip=0} \vspace{11.5cm} \caption{\it
Transition probability $P(\nu_e \rightarrow \nu_\mu)$ as a function of
the baseline length $L$. Here, $\varepsilon_{e\tau}=0.01$ and
$\delta=\pi/2$ are adopted, and the neutrino beam energies have been
labeled in each plot. The solid and dotted curves denote the exact
numerical results with and without NSIs, respectively. Probabilities
calculated using our approximate mappings are shown as dashed curves.}
\label{fig:fig5} \vspace{0. cm}
\end{center}
\end{figure}

\begin{figure}[t]
\begin{center}
\vspace{-4.5cm}
\epsfig{file=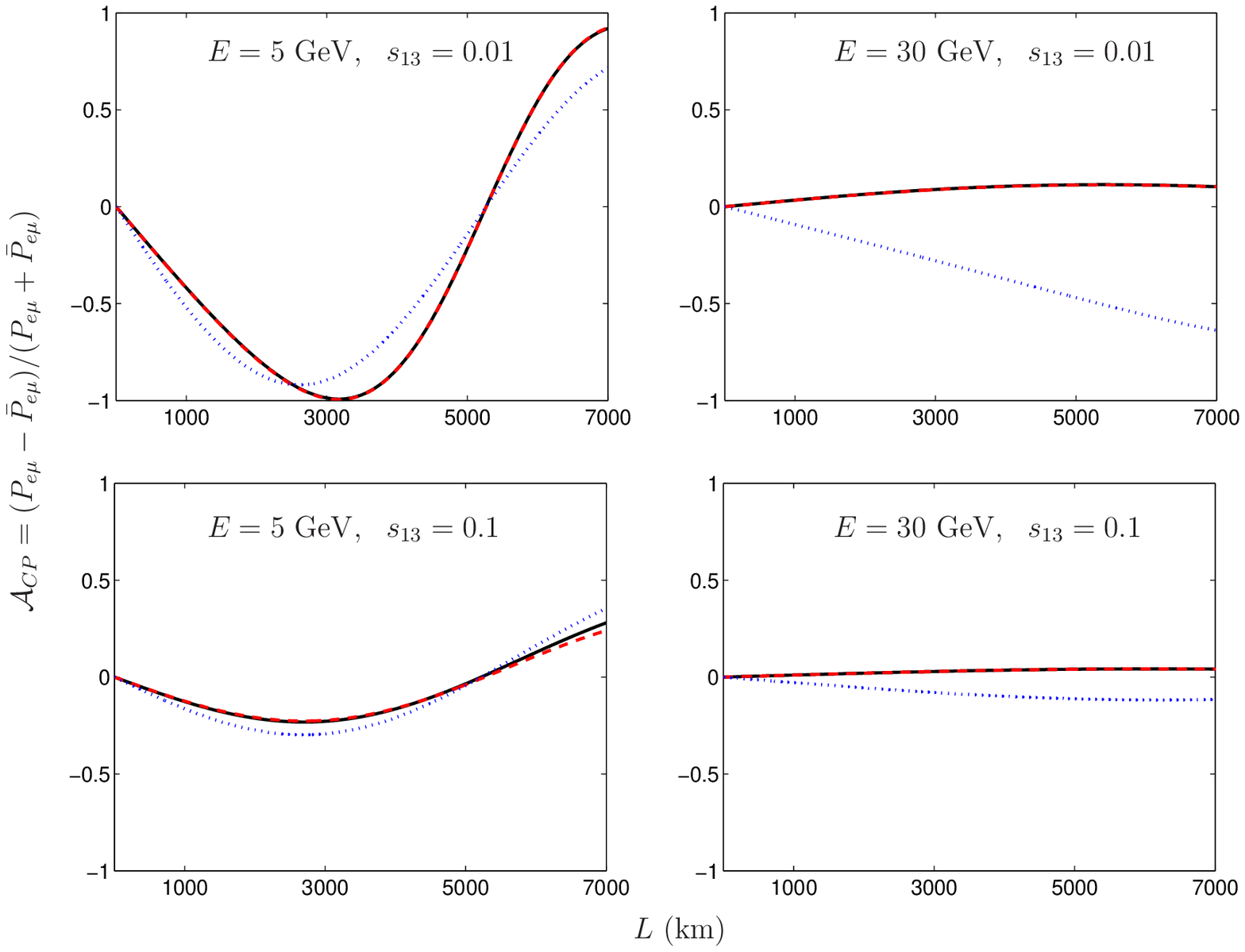,bbllx=9cm,bblly=23cm,bburx=13cm,bbury=27cm,%
width=3.cm,height=3.cm,angle=0,clip=0} \vspace{11.5cm} \caption{\it
CP asymmetries as a function of the baseline lengths. Here,
$\varepsilon_{e\tau}=0.01$ and the neutrino beam energies have been
shown in each plot. The solid and dotted curves denote the exact
numerical results with and without NSIs, respectively. Dashed curves
denote ${\cal A}_{CP}$ calculated using the approximate mappings.}
\label{fig:fig6} \vspace{0. cm}
\end{center}
\end{figure}

In Figs.~\ref{fig:fig5} and \ref{fig:fig6}, we repeat the same
exercise on the neutrino oscillation probabilities and CP
asymmetries, but instead as a function of the baseline length and
for two fixed value of the neutrino energy $E=5$ GeV and $E=30$ GeV.
%%%%%%%%%%%%%%%%%%%%%%%%%%%%%%%%%%%%%%%%%%%
%%%%%%%%%%%%%%%%%%%% Fig.~5 %%%%%%%%%%%%%%%%
%%%%%%%%%%%%%%%%%%%%%%%%%%%%%%%%%%%%%%%%%%%
%%%%%%%%%%%%%%%%%%%%%%%%%%%%%%%%%%%%%%%%%%%
%%%%%%%%%%%%%%%%%%%%%%%%%%%%%%%%%%%%%%%%%%%
%%%%%%%%%%%%%%%%%%%% Fig.~6 %%%%%%%%%%%%%%%%
%%%%%%%%%%%%%%%%%%%%%%%%%%%%%%%%%%%%%%%%%%%
%%%%%%%%%%%%%%%%%%%%%%%%%%%%%%%%%%%%%%%%%%%
It can be clearly seen that for smaller $\theta_{13}$ and {lower}
beam energy, new physics effects play a significant role around
$L\sim 3000~{\rm km}$, which sheds some light on future beta beam
experiments. For higher energy experiments, i.e., a neutrino
factory, a far detector with relatively longer baseline length
should be important to constrain NSI effects. In both figures, one
can appreciate how the probabilities computed using our
approximations for the effective mixing angles are in very good
agreement with the exact results.

\begin{figure}[t]
\begin{center}
\vspace{-4.5cm}
\epsfig{file=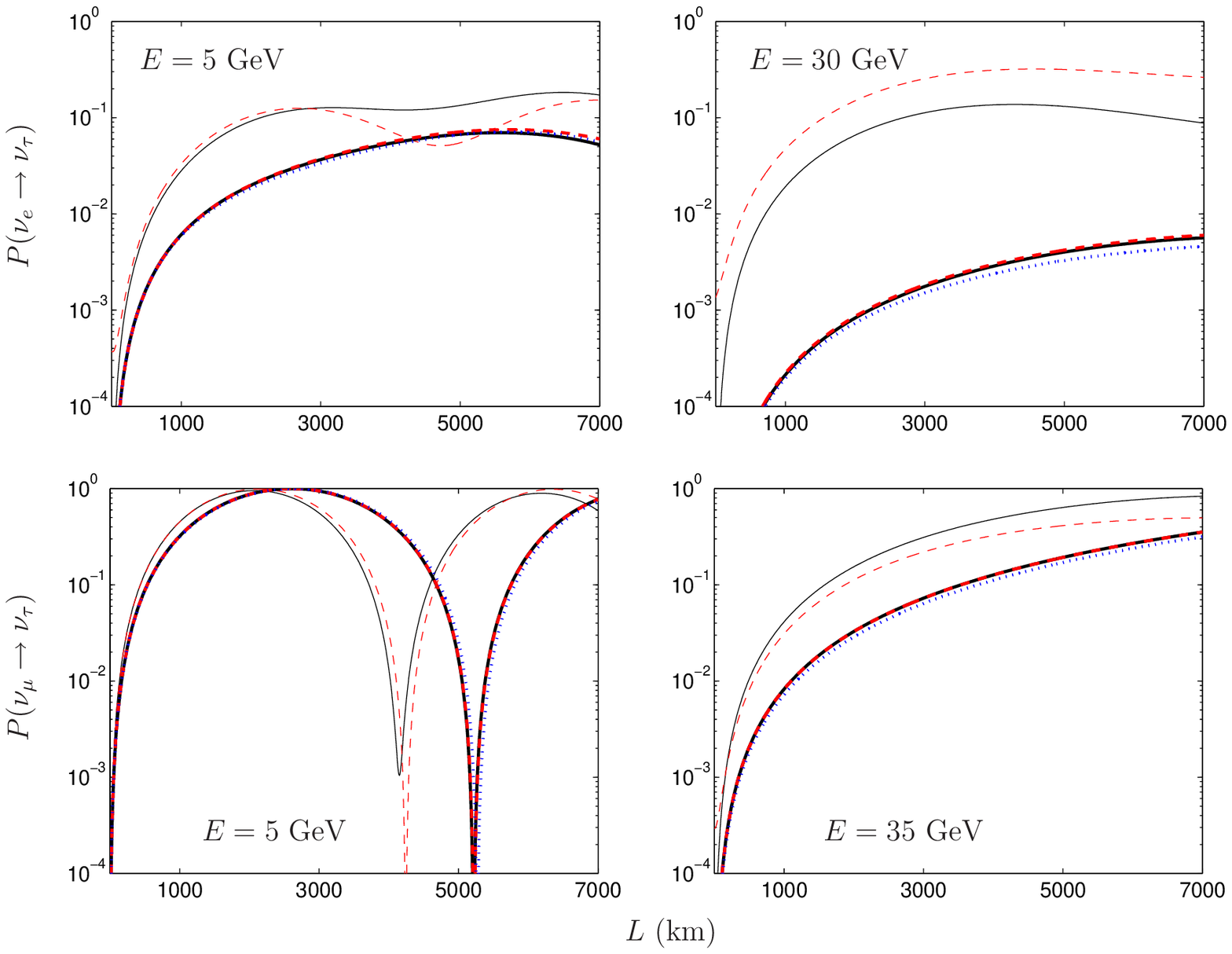,bbllx=9cm,bblly=23cm,bburx=13cm,bbury=27cm,%
width=3.cm,height=3.cm,angle=0,clip=0} \vspace{11.5cm} \caption{\it
Transition probabilities for the $\nu_e \rightarrow \nu_\tau$ and
$\nu_\mu \rightarrow \nu_\tau$ channels as a function of the
baseline length $L$. Different setups of NSI parameters are
considered: (a) for the $\nu_e \rightarrow \nu_\tau$ channel, we
choose $\varepsilon_{e\tau} = 0.01$, and for the $\nu_\mu
\rightarrow \nu_\tau$ channel, we choose and $\varepsilon_{\mu\mu} =
\varepsilon_{\mu\tau} = \varepsilon_{\tau\tau} = 0.01$. Here, thick
solid and dashed curves correspond to the exact numerical results
and our approximate mappings, respectively; (b) NSI parameters are
at their upper bounds computed in Ref.~\cite{Ribeiro:2007ud} with
thin solid and dashed curves corresponding to the exact numerical
results and our approximate mappings, respectively. Dotted curves
denote the numerical results without including NSIs, and they are
unique in each channel.} \label{fig:fig7} \vspace{0. cm}
\end{center}
\end{figure}

Finally, we illustrate the application of our analytical expressions
for the $\nu_e \rightarrow \nu_\tau$ and $\nu_\mu \rightarrow
\nu_\tau$ channels in Fig.~\ref{fig:fig7}.
%%%%%%%%%%%%%%%%%%%%%%%%%%%%%%%%%%%%%%%%%%%
%%%%%%%%%%%%%%%%%%%% Fig.~7 %%%%%%%%%%%%%%%%
%%%%%%%%%%%%%%%%%%%%%%%%%%%%%%%%%%%%%%%%%%%
%%%%%%%%%%%%%%%%%%%%%%%%%%%%%%%%%%%%%%%%%%%
For comparison, we also show the maximal NSI corrections by setting
all the NSI parameters at their upper bounds given in
Ref.~\cite{Ribeiro:2007ud}. One can observe that NSI corrections to
these two channels are not remarkable if the corresponding
$\varepsilon_{\alpha\beta}$'s are chosen to be a few percent.
However, increasing the NSI parameters, the NSI effects become more
significant, in particular for the $\nu_e \rightarrow \nu_\tau$
channel. The upper plots in Fig.~\ref{fig:fig7} indicate that our
approximate mappings are not quite valid for relatively longer
baseline lengths. This is due to the fact that our expansions are
performed according to small $\varepsilon_{\alpha \beta}$'s and
cannot be extended to the regions of sizable NSI parameters. As
discussed in the introduction, if NSIs are related to some
underlying new physics, they should be attributed to next-to-leading
order effects and not deviate much from zero. In this sense, our
approximate mappings are quite realistic and should be very helpful
for both phenomenological studies and model buildings.

Since the analyses above certainly depend on the input NSI parameters,
they mainly serve as illustrations. However, our analytical results
are model independent. Thus, they are hoped to be very useful for a
general study of NSI effects in future experiments. The transparent
mappings also manifest the underlying correlations between leptonic
mixing parameters and NSIs in a very legible way.

\section{Summary}
\label{sec:summary}

In this work, we have developed both exact and approximate mappings
between the leptonic mixing matrix in vacuum and in matter in the
presence of NSIs. A full set of sum rules between fundamental mixing
parameters and the corresponding effective ones in matter have been
derived. By using these sum rules, exact and model independent
analytical mappings between the mixing matrix elements $\tilde
V_{\alpha i}$ and $V_{\alpha i}$ have been established, and in turn
using these mappings, the moduli of the mixing matrix elements and
the sides of unitarity triangles can be immediately figured out.
Besides the exact expressions for the mixing parameters, we have
also derived approximate parameter mappings based on series
expansions in the small parameters $\eta$, $s_{13}$, and
$\varepsilon_{\alpha \beta}$. We have then performed a detailed
numerical analysis of the application and validity of our parameter
mappings. In particular, we have concentrated on the mimicking
effects of NSIs on the mixing angle $\theta_{13}$ and on the
deviation of the mixing angle $\theta_{23}$ from maximal mixing.
Furthermore, we have studied in detail how the $\varepsilon_{\alpha
\beta}$'s affect the transition probabilities of the $\nu_e
\rightarrow \nu_{\mu}$, $\nu_e \rightarrow \nu_{\tau}$, and $\nu_\mu
\rightarrow \nu_\tau$ channels. We have found that the exact
parameter mappings are very useful in obtaining exact results for
the mixing parameters and transition probabilities, and our
perturbative parameter mappings also describe quite well all the
relevant features of these quantities. Note that our analytical
analysis is independent of any specific model or assumptions on the
configuration of NSI parameters. In conclusion, the outstanding
feature of our parameter mappings is that they reveal the underlying
correlations between NSI effects and neutrino mixing parameters in a
highly straightforward way, and they are very practical and useful
for the study of NSIs in future long-baseline neutrino oscillation
experiments. It also makes sense to note that the calculation
procedures we have employed in the present work can also be applied
to the picture of non-unitary leptonic mixing
\cite{Antusch:2006vwa}, which will be elaborated on elsewhere.

\begin{acknowledgments}
We would like to thank Mattias Blennow for useful discussions.
This work was supported by the Royal Swedish Academy of Sciences (KVA)
[T.O.], the G{\"o}ran Gustafsson Foundation [H.Z.], and the Swedish
Research Council (Vetenskapsr{\aa}det), contract no.~621-2008-4210
[T.O.]. D.M. acknowledges partial financial support from the Ministry
of University and Scientific Research of Italy, through the 2007-08
COFIN program.
\end{acknowledgments}

\newpage

\appendix
\section{Calculations of effective masses}
\label{app:effective_masses}

To calculate the explicit expressions of $\tilde{m}_i$, the cubic
roots of the characteristic polynomial of Eq.~\eqref{eq:Hm} are
involved. We follow the method given in Ref.~\cite{Kopp:2008} and
define the so-called elementary symmetric polynomials
\cite{Ohlsson:1999xb,Ohlsson:1999um}:
\begin{eqnarray}\label{eq:A1}
c_0 & = & \tilde{H}_{ee}\left|\tilde{H}_{\mu \tau}\right|^2 +
\tilde{H}_{\mu \mu}\left|\tilde{H}_{e \tau}\right|^2 +
\tilde{H}_{\tau \tau}\left|\tilde{H}_{e \mu}\right|^2 - 2{\rm Re}
(\tilde{H}_{e \mu} \tilde{H}_{\mu \tau}\tilde{H}_{\tau e}) -
\tilde{H}_{ee}\tilde{H}_{\mu \mu}\tilde{H}_{\tau \tau} \ ,
\\ \label{eq:A2}
c_1 & = &  \tilde{H}_{ee}\tilde{H}_{\mu \mu} +
\tilde{H}_{ee}\tilde{H}_{\tau \tau}  + \tilde{H}_{\mu
\mu}\tilde{H}_{\tau \tau} - \left|\tilde{H}_{e \mu}\right|^2 -
\left|\tilde{H}_{\mu \tau}\right|^2 -
\left|\tilde{H}_{e \tau}\right|^2 \ , \\
c_2 & = & - \tilde{H}_{ee} - \tilde{H}_{\mu \mu} - \tilde{H}_{\tau
\tau} \ . \label{eq:A3}
\end{eqnarray}
It is easy to check that the relations $c_2 = -\sum_i
\tilde{m}^2_i/(2E)$, $c_1 = \sum_{i<j} \tilde{m}^2_i
\tilde{m}^2_j/(2E)^2$, and $c_0=-\prod_i \tilde{m}^2_i/(2E)^3$ are
satisfied. By incorporating the definitions above, the mass squared
eigenvalues can be computed as
\begin{eqnarray}\label{eq:A4}
\frac{\tilde{m}^2_1}{2E} & = & \frac{2}{3}\sqrt{p} \cos
\left[\frac{1}{3}\arctan \left(\frac{\sqrt{p^3-q^2}}{q} \right)+
\frac{2\pi}{3}\right] -\frac{1}{3} c_2 \  ,  \\
\label{eq:A5} \frac{\tilde{m}^2_2}{2E} & = &  \frac{2}{3}\sqrt{p}
\cos \left[\frac{1}{3}\arctan \left(\frac{\sqrt{p^3-q^2}}{q}\right)
-
\frac{2\pi}{3}\right] -\frac{1}{3} c_2 \  ,  \\
\frac{\tilde{m}^2_3}{2E} &= & \frac{2}{3}\sqrt{p} \cos
\left[\frac{1}{3}\arctan \left( \frac{\sqrt{p^3-q^2}}{q} \right)
\right] -\frac{1}{3} c_2 \ , \label{eq:A6}
\end{eqnarray}
where $p=c^2_2-3c_1$ and $q= -27 c_0/2 -c^3_2+9 c_1 c_2/2$. As a
natural consequence, the effective mass eigenvalues in matter are only
related with the neutrino mass squared differences but not the
absolute neutrino masses.

As an example, we consider the case of vanishing NSI. From
Eqs.~\eqref{eq:A1}-\eqref{eq:A3}, one can directly write down
\begin{eqnarray}\label{eq:A7}
c_0 & = & - \frac{1}{(2E)^3} A \Delta_{21} \Delta_{31}
\left|V_{e1}\right|^2 \ ,  \\ \label{eq:A8} c_1 & = &
\frac{1}{(2E)^2} \left\{ \Delta_{21} \Delta_{31} + A \left[
\Delta_{21}\left(1-\left|V_{e2}\right|^2\right) + \Delta_{31}
\left(1-\left|V_{e3}\right|^2\right) \right] \right\} \ ,  \\
c_2 & = & - \frac{1}{2E} \left( A + \Delta_{21} + \Delta_{31}
\right)\ . \label{eq:A9}
\end{eqnarray}
Substituting Eqs.~\eqref{eq:A7}-\eqref{eq:A9} into
Eqs.~\eqref{eq:A4}-\eqref{eq:A6}, the matter corrected eigenvalues
given in Refs.~\cite{Barger:1980tf,Zaglauer:1988gz} can be
reproduced straightforwardly. One may also check that $A=0$ leads to
the limit $\tilde{m}_i = m_i$.

\section{Formulas for Effective Mixing Matrix Elements}
\label{app:matrix elements}

In the limit of small parameters, i.e., $\eta \to 0$ and $V_{e3} \to
0$, one can use our main result for the exact analytical parameter
mappings Eq.~\eqref{eq:solution} to derive zeroth-order series
expansion formulas for the modulus squares of the mixing matrix
elements $V_{e3}$, $V_{e2}$, and $V_{\mu 3}$. The results are given by
\begin{eqnarray}
|\tilde V_{e3}|^2 &=& \frac{1}{\tilde\Delta_{31} \tilde\Delta_{32}}
\left\{ \tilde m_1^2 \tilde m_2^2 - A (1 + \varepsilon_{ee}) (\tilde
m_1^2 + \tilde m_2^2) + A^2 \left[ (1 + \varepsilon_{ee})^2 +
  |\varepsilon_{e \mu}|^2 +
|\varepsilon_{e \tau}|^2 \right] \right\} \ , \nonumber \\ \\
|\tilde V_{e2}|^2 &=& \frac{1}{\tilde\Delta_{21} \tilde\Delta_{23}}
\left\{ \tilde m_1^2 \tilde m_3^2 - A (1 + \varepsilon_{ee}) (\tilde
m_1^2 + \tilde m_3^2) + A^2 \left[ (1 + \varepsilon_{ee})^2 +
|\varepsilon_{e \mu}|^2 + |\varepsilon_{e\tau}|^2 \right] \right\} \
, \nonumber \\ \\ |\tilde V_{\mu 3}|^2 &=&
\frac{1}{\tilde\Delta_{31} \tilde\Delta_{32}} \left\{ \tilde m_1^2
\tilde m_2^2 + \Delta_{31} \left( \Delta_{31} - \tilde m_1^2 -
\tilde m_2^2 \right) |V_{\mu 3}|^2 + A^2 \left( |\varepsilon_{e
\mu}|^2 + |\varepsilon_{\mu\mu}|^2 + |\varepsilon_{\mu\tau}|^2
\right) \right. \nonumber \\ && \left. - A \varepsilon_{\mu\mu}
\left( \tilde m_1^2 + \tilde m_2^2 \right) + 2 A \Delta_{31} \left[
\varepsilon_{\mu\mu} |V_{\mu 3}|^2 + {\rm Re}(\varepsilon_{e\mu}
V_{e 3} V^*_{\mu 3}) + {\rm Re}(\varepsilon_{\mu\tau} V_{\tau 3}
V^*_{\mu 3}) \right] \right\} \ , \nonumber \\
\end{eqnarray}
which are valid to all orders in the NSI parameters. In addition,
for standard matter effects, i.e., without NSI effects, and for any
$\eta$ and $V_{e3}$, we can derive the corresponding formula to
Eq.~\eqref{eq:solution}. The result is
\begin{eqnarray}
\tilde V_{\alpha i} \tilde V^*_{\beta i} & = & \frac{1}{\tilde
\Delta_{im} \tilde \Delta_{in}} \left[ \sum_j \hat\Delta_{jm}
\hat\Delta_{jn} V_{\alpha j} V^*_{\beta j} + A \delta_{\alpha e}
\delta_{\beta e} \left( A - \tilde m_n^2 - \tilde m_m^2 \right)
\right. \nonumber \\ && \left. ~~~~~~~~~~~~~~~ + A \sum_j
\Delta_{j1} \left( \delta_{\alpha e} V_{e j} V^*_{\beta j} +
\delta_{\beta e} V_{\alpha j} V^*_{e j} \right) \right] \ .
\end{eqnarray}
In the specific cases of $V_{e3}$, $V_{e2}$, and $V_{\mu 3}$, we
obtain
\begin{eqnarray}
|\tilde V_{e3}|^2 &=& \frac{1}{\tilde\Delta_{31} \tilde\Delta_{32}}
\left[ \sum_j \hat\Delta_{j1} \hat\Delta_{j2} |V_{ej}|^2 + A \left(
A - \tilde m_1^2 - \tilde m_2^2 \right) + 2 A \sum_{j = 2,3}
\Delta_{j1} |V_{ej}|^2 \right] \ , \\
|\tilde V_{e2}|^2 &=& \frac{1}{\tilde\Delta_{21} \tilde\Delta_{23}}
\left[ \sum_j \hat\Delta_{j1} \hat\Delta_{j3} |V_{ej}|^2 + A \left(
A - \tilde m_1^2 - \tilde m_3^2 \right) + 2 A \sum_{j = 2,3}
\Delta_{j1} |V_{ej}|^2 \right] \ , \\ |\tilde V_{\mu 3}|^2 &=&
\frac{1}{\tilde\Delta_{31} \tilde\Delta_{32}} \sum_j \hat\Delta_{j1}
\hat\Delta_{j2} |V_{\mu j}|^2 \ ,
\end{eqnarray}
which are valid to all orders in the small parameters $\eta$ and
$V_{e3}$.

%\bibliography{NSI}

\end{document}